%% file: TOP-24-011_temp.tex
\documentclass[11pt,twoside,a4paper,cmspaper,final,collab]{cms-tdr}
\def\svnVersion{debbfef}\def\svnDate{2026/08/29}\def\cmsCernNoTag{CERN-EP-2025-289}\def\cmsCernDate{\today}\def\cmsMessage{Published in the Journal of High Energy Physics as \href{http://dx.doi.org/10.1007/JHEP08(2026)223}{\doi{10.1007/JHEP08(2026)223}.}}
\begin{document}\cmsNoteHeader{TOP-24-011}

\providecommand{\cmsTable}[1]{\resizebox{\textwidth}{!}{#1}}

\newcommand{\mtop}{\ensuremath{m_{\text{top}}}\xspace}
\newcommand{\mtopreco}{\ensuremath{m_{\text{top}}^{\text{reco}}}\xspace}
\newcommand{\mtw}{\ensuremath{m_\mathrm{T}^{\PW}}\xspace}
\newcommand{\tq}{\ensuremath{\PQt\PQq}\xspace}
\newcommand{\tbarq}{\ensuremath{\PAQt\PQq}\xspace}
\newcommand{\tqandtbarq}{\ensuremath{\tq\text{+}\tbarq}\xspace}
\newcommand{\tW}{\ensuremath{\PQt\PW}\xspace}
\newcommand{\tch}{\ensuremath{t\text{ channel}}\xspace}

\newcommand{\wjets}{\ensuremath{\PW\text{+jets}}\xspace}
\newcommand{\wlight}{\ensuremath{\PW\text{+light}}\xspace}
\newcommand{\wc}{\ensuremath{\PW\text{+}\PQc}\xspace}
\newcommand{\wb}{\ensuremath{\PW\text{+}\PQb}\xspace}
\newcommand{\mW}{\ensuremath{m_{\PW}}\xspace}
\newcommand{\pb}{\unit{pb}}

\newcommand{\asmz}{\ensuremath{\alpS(m_{\PZ})}\xspace}
\newcommand{\mur}{\ensuremath{\mu_{\mathrm{R}}}\xspace}
\newcommand{\muf}{\ensuremath{\mu_{\mathrm{F}}}\xspace}
\newcommand{\abseta}{\ensuremath{\abs{\eta}}\xspace}

\newcommand{\snorm}{\ensuremath{\sigma_{\text{norm}}}\xspace}
\newcommand{\pp}{\ensuremath{\Pp\Pp}\xspace}
\newcommand{\sto}{\ensuremath{\sigma(\tq)}\xspace}
\newcommand{\sboth}{\ensuremath{\sigma(\tqandtbarq)}\xspace}
\newcommand{\santitop}{\ensuremath{\sigma(\tbarq)}\xspace}
\newcommand{\hdamp}{\ensuremath{h_{\text{damp}}}\xspace}
\newcommand{\ejets}{\ensuremath{\Pe\text{+jets}}\xspace}
\newcommand{\mjets}{\ensuremath{\PGm\text{+jets}}\xspace}
\newcommand{\njnb}[2]{\ensuremath{#1\PQj#2\PQb}\xspace}

\newcommand{\PQj}{\ensuremath{\HepParticle{j}{}{}}\xspace}
\newcommand{\PQjpr}{\ensuremath{\HepParticle{j}{}{\prime}}\xspace}

\newcommand{\httprime}{\ensuremath{H_{\mathrm{T}}^\prime}\xspace}
\newcommand{\deltarmed}{\ensuremath{\DR_{\text{med}}(\PQj,\PQjpr)}\xspace}
\newcommand{\mjj}{\ensuremath{m(\PQj,\PQjpr)}\xspace}
\newcommand{\mlb}{\ensuremath{m(\Pell,\PQb)}\xspace}
\newcommand{\absueta}{\ensuremath{\abs{\eta_{\PQu_0}}}\xspace}
\newcommand{\deltaeta}{\ensuremath{\abs{\Delta\eta(\PQu,\PQb)}}\xspace}
\newcommand{\sumvecall}{\ensuremath{\HT(\PQj,\Pell)}\xspace}
\newcommand{\dptwub}{\ensuremath{\abs{\Delta\pt(\PW,\PQu\PQb)}}\xspace}
\newcommand{\dphiwub}{\ensuremath{\abs{\Delta\phi(\PW,\PQu\PQb)}}\xspace}
\newcommand{\detalu}{\ensuremath{\abs{\Delta\eta(\Pell,\PQu)}}\xspace}

\newcommand{\nSR}[1]{\ensuremath{N^{\text{SR}}_{\text{#1}}}\xspace}
\newcommand{\nCR}[1]{\ensuremath{N^{\text{CR}}_{\text{#1}}}\xspace}
\newcommand{\Irel}{\ensuremath{I_{\text{rel}}}\xspace}

\newcommand{\DeepCSV}{\ensuremath{\textsc{DeepCSV}}\xspace}
\newcommand{\fLV}{\ensuremath{f_{\text{LV}}}\xspace}
\newcommand{\Vtb}{\ensuremath{V_{\PQt\PQb}}\xspace}
\newcommand{\absfLVVtb}{\ensuremath{\abs{\fLV\Vtb}}\xspace}
\newcommand{\experr}{\ensuremath{\,\text{(exp)}}\xspace}
\newcommand{\ratio}{\ensuremath{\mathcal{R}_{\mathrm{t-ch}}\xspace}}

\cmsNoteHeader{TOP-24-011}
\title{Measurement of the \texorpdfstring{$t$}{t}-channel single top quark cross section in proton-proton collisions at \texorpdfstring{$\sqrt{s}=5.02\TeV$}{sqrt(s)=5.02 TeV}}

\date{\today}

\abstract{The single top quark $t$-channel production cross section is measured in proton-proton collisions at the CERN LHC at $\sqrt{s}=5.02\TeV$, using data recorded with the CMS detector in 2017,
corresponding to an integrated luminosity of 302\pbinv, and resulting in the first CMS measurement of the process at that energy. Events with one electron or muon and two or more jets,
among which at least one is identified as originating from a \PQb quark fragmentation, are analyzed. The combined cross section of single top quark (\tq) and single top antiquark (\tbarq) production is $\sboth=25.5\,^{+3.6}_{-3.5}\stat\,^{+4.2}_{-3.9}\syst\pm0.5\lum\pb$. The individual cross sections are measured
to be $\sto=17.7\,^{+2.9}_{-2.7}\stat\,^{+2.6}_{-2.4}\syst\pm0.3\lum\pb$
and $\santitop=6.7\,^{+2.4}_{-1.6}\stat\,^{+2.1}_{-2.5}\syst\pm0.1\lum\pb$.
Their ratio is measured to be $\ratio=2.7\,^{+1.5}_{-0.8}\stat\,^{+1.2}_{-0.3}\syst$.
The absolute value of the Cabibbo--Kobayashi--Maskawa matrix element is found to be $\absfLVVtb=0.92\pm0.09\experr\pm0.01\thy$.
The measurements are in good agreement with the standard model predictions at next-to-next-to-leading order accuracy in quantum chromodynamics.}

\hypersetup{%
pdfauthor={CMS Collaboration},%
pdftitle={Measurement of the t-channel single top quark cross section in proton-proton collisions at sqrt(s)=5.02 TeV},%
pdfsubject={CMS},%
pdfkeywords={CMS, top quark}}

\maketitle

\section{Introduction}
\label{sec:intro}

\begin{figure}[!ht]
\centering
\includegraphics[width=0.4\textwidth]{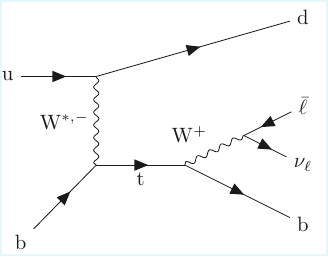}
\hspace{0.1\textwidth}
\includegraphics[width=0.4\textwidth]{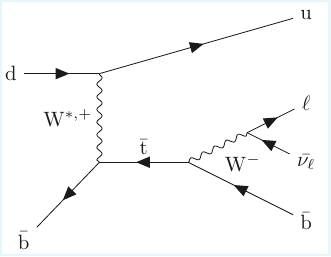}
\caption{Representative leading-order Feynman diagrams for single top quark (left) and antiquark (right) production via the $t$ channel.}
\label{fig:feynman}
\end{figure}

The top quark is the heaviest fundamental particle discovered so far. Because of its large mass and decay width, it has a very short lifetime of about $10^{-25}\unit{s}$~\cite{Bigi:1986jk}, an order of magnitude shorter than the characteristic timescale on which hadron bound states form.
As a consequence, it is the only quark that can be studied as an asymptotically free particle. At the LHC, single top quark production in proton-proton (\pp) collisions proceeds dominantly via the $t$-channel process, yielding polarized top quarks. Single top quarks in the $t$ channel are produced with the exchange of a \PW boson through an initial-state bottom (\PQb) quark. This results in sensitivity to the \PQb quark parton distribution function (PDF),
in particular in the top quark to antiquark cross section ratio. The $t$ channel has a characteristic final-state topology with the presence of a light quark recoiling against the top quark. Figure~\ref{fig:feynman} shows representative Feynman diagrams of $t$-channel single top quark and antiquark production.
The single top quark production cross section is proportional to the squared Cabibbo--Kobayashi--Maskawa matrix element $\abs{\Vtb}$ , allowing for a direct measurement of this quantity to further test the standard model (SM), including the assumption that only
three quark families exist~\cite{theo:Alwall_2006,theo:Holdom_2009}.

A measurement of single top quark production in the $t$ channel was performed for the first time at the Tevatron collider at Fermilab~\cite{Tevatron}. More recently, both the ATLAS and CMS Collaborations have reported measurements of the cross section for single top quark production in the $t$ channel using LHC data collected at $\sqrt{s}=7$, 8, and 13\TeV~\cite{Chatrchyan:2011vp,Chatrchyan:2012ep,atlas7,Khachatryan:2014iya,atlas8,Sirunyan:2016cdg,top201711,atlas13,ATLAS:2026vxj}, both inclusively and differentially, and with increased precision.
The ATLAS Collaboration has measured the process at $\sqrt{s}=5.02\TeV$ using data corresponding to an integrated luminosity of 255\pbinv, obtaining a cross section of $\sboth=27.1\,^{+6.2}_{-5.5}\pb$, in agreement with the SM prediction~\cite{atlas5tev}.

This paper presents the first CMS measurement of the single top quark production cross section at $\sqrt{s}=5.02\TeV$, using data collected by CMS during a special LHC run in 2017, corresponding to an integrated luminosity of 302\pbinv. Data recorded in the same run have been analyzed by CMS to perform several measurements, including cross sections of diboson~\cite{wz5tev} and top quark pair (\ttbar)~\cite{CMS:2021gwv,CMS:2024ghc} production.
Performing measurements at various beam energies provides interesting stress tests for the theory, probing whether current calculations correctly describe the energy dependence of the measured processes.
Moreover, in the study of the \PQb quark PDF in the proton, complementary insight can be gained by probing different regions of the proton momentum fraction.
A distinct feature of this data sample is the low number of interactions per bunch crossing (pileup) with respect to the standard operating conditions of the LHC, yielding a cleaner detector environment, and improved detection and reconstruction.
In this measurement, the top quark is reconstructed via the leptonic decay channel of the \PW boson, $\PW\to\Pell\PGn$ (with $\Pell=\Pe,\PGm$), since it provides a cleaner experimental signature when compared to the fully hadronic \PW boson decay channel.
The measured cross sections are obtained by performing binned maximum likelihood (ML) fits of histogram templates to the data in several event categories.

This paper is organized as follows. A brief description of the CMS detector is given in Section~\ref{sec:CMS}.
Section~\ref{sec:dataandMC} describes the data used for the measurement, as well as the Monte Carlo (MC) simulation samples,
followed by the object reconstruction and event selection in Section~\ref{sec:EventSelection}. Background estimation methods are covered in
Section~\ref{sec:Background}, while systematic uncertainties are described
in Section~\ref{sec:Systematics}. Results are discussed in Section~\ref{sec:Results}, and the summary is given in Section~\ref{sec:Summary}.
Tabulated results from this analysis are provided in the HEPData record~\cite{hepdata}.

\section{The CMS detector and object reconstruction}
\label{sec:CMS}

The CMS apparatus~\cite{CMS:2008zzk,CMS:2023gfb} is a multipurpose, nearly hermetic detector, designed to trigger on~\cite{Sirunyan:2020zal,Khachatryan:2016bia,CMS:2024aqx}
and identify electrons, muons, photons, charged and neutral hadrons~\cite{CMS:2020uim,Sirunyan:2018fpa,CMS:2014pgm}.
Its central feature is a superconducting solenoid of 6\unit{m} internal diameter, providing a magnetic field of 3.8\unit{T}.
Within the solenoid volume are a silicon pixel and strip tracker, a lead tungstate crystal electromagnetic calorimeter (ECAL),
and a brass and scintillator hadron calorimeter (HCAL), each composed of a barrel and two endcap sections.
Forward calorimeters extend the pseudorapidity ($\eta$) coverage provided by the barrel and endcap detectors.
Muons are reconstructed using gas-ionization detectors interleaved with the layers of the steel flux-return yoke outside the solenoid.
More detailed descriptions of the CMS detector, together with a definition of the coordinate system used and the relevant kinematic variables,
can be found in Refs.~\cite{CMS:2008zzk,CMS:2023gfb}.

Events of interest are selected using a two-tiered trigger system. The first level (L1),
composed of custom hardware processors, uses information from the calorimeters and muon detectors to select events
at a rate of around 100\unit{kHz} within a fixed latency of about 4\mus~\cite{Sirunyan:2020zal}. The second level, known as the high-level
trigger (HLT), consists of a farm of processors running a version of the full event reconstruction software optimized
for fast processing, and reduces the event rate to around 30\unit{kHz} in this special run before data storage~\cite{Khachatryan:2016bia,CMS:2024aqx}. We consider in the analysis those events in which the trigger detected at least one electron (muon) with transverse momentum $\pt>20$ (17)\GeV.

The particle-flow (PF) algorithm~\cite{CMS-PRF-14-001} aims to reconstruct and identify each individual particle in an event, with an optimized combination of information from the various elements of the CMS detector. The energy of photons is obtained from the ECAL measurement. The energy of electrons is determined from a combination of the electron momentum at the primary interaction vertex (PV) as determined by the tracker, the energy of the corresponding ECAL cluster, and the energy sum of all bremsstrahlung photons spatially compatible with originating from the electron track. The energy of muons is obtained from the curvature of the corresponding track.
The energy of charged hadrons is determined from a combination of their momentum measured in the tracker and the associated energy deposits in the ECAL and HCAL, corrected for the calorimeter response to hadronic showers. In addition, the energy of neutral hadrons is obtained from the corresponding corrected ECAL and HCAL energies. Finally, the missing transverse momentum \ptvecmiss is defined as the negative vector \pt sum of all PF candidates~\cite{ptmissperformance}.

Jets are clustered from all objects reconstructed by the PF algorithm using the anti-\kt algorithm~\cite{Cacciari:2008gp, Cacciari:2011ma} with a distance parameter of 0.4. The jet momentum is determined as the vectorial sum of all particle momenta in the jet, and is found from simulation to be, on average, within 5--10\% of the true momentum over the whole \pt spectrum and detector acceptance. Jet energy corrections are derived from simulation studies so that the average measured energy of jets becomes identical to that of particle-level jets. Simulated particle-level jets are obtained by clustering stable particles originating from the hadronization of final-state partons. To determine any residual differences between the jet energy scale (JES) in data and in simulation, measurements of the momentum balance are used, leading to appropriate correction factors. Additional selection criteria are applied to each jet to remove jets potentially dominated by instrumental effects or reconstruction failures. This is done by rederiving the corrections for this specific run with the methods used in Refs.~\cite{Khachatryan:2016kdb,CMS:2017wyc}. Jets are identified as coming from the fragmentation of bottom quarks using the medium working point of the \DeepCSV \PQb tagging algorithm~\cite{Sirunyan:2017ezt}, calibrated at 13\TeV, as differences with the 5.02\TeV data set are negligible. Nevertheless, the corresponding efficiencies are derived for the latter dataset. For jets with $\pt>20\GeV$ in \ttbar events we obtain a \PQb tagging efficiency of 68\%, and mistag rates of 12\% and 1.1\% for \PQc quark and light-flavour quark or gluon jets, respectively. We will refer to jets in general as ``\PQj'', to \PQb-tagged jets as ``\PQb'', and to untagged (non-\PQb-tagged) jets as ``\PQu''.

A gradient boosted decision tree (BDT) based on the shower shape and the track quality of the electron candidates~\cite{CMS:2020uim} is used to identify electrons. Electron candidates are required to satisfy $\abseta<2.4$ (excluding the region $1.48<\abseta<1.57$, the transition region between the barrel and endcaps, where the electron reconstruction is suboptimal) and $\pt>20\GeV$. A veto is applied on candidates that are matched to a secondary vertex consistent with a photon conversion, or have a missing hit in the inner layer of the tracker. Reconstructed muon candidates are required to have $\abseta<2.4$ and $\pt>20\GeV$. They must also fulfil criteria on the
geometrical matching between the tracks reconstructed by the silicon tracker and the muon system, as well as on the quality of the global muon track fit combining information from both subdetectors~\cite{Sirunyan:2018fpa}. Both electron and muon energy scales and resolution are corrected according to measurements using Drell--Yan (DY) events in simulation and low-pileup \pp collision data at 5.02\TeV~\cite{CMS:2020uim,Sirunyan:2018fpa}.

Pileup collisions may cause events to contain multiple interaction vertices. The PV is taken to be the vertex corresponding to the hardest scattering in the event, which is evaluated using tracking information alone, as described in Section~9.4.1 of Ref.~\cite{CMS-TDR-15-02}.

Lepton candidates must be consistent with originating from the PV. This is guaranteed by requiring the longitudinal impact parameter to be smaller or equal to 0.1\unit{cm} and the transverse impact parameter smaller than 0.02 (0.05)\unit{cm} for electrons (muons).
Electrons and muons must also satisfy
a requirement on their relative isolation (\Irel), defined as the scalar \pt sum of all the particles inside a cone around the lepton direction, excluding the lepton itself, divided by the lepton \pt.
The cone size, defined as $\DR=\sqrt{\smash[b]{(\Delta\eta)^2+(\Delta\phi)^2}}$, where $\phi$ is the azimuthal angle, changes as a function of the lepton \pt as $\DR(\pt)=10\GeV/\pt$ if $50<\pt<200\GeV$, $\DR=0.2$ if $\pt<50\GeV$, and $\DR=0.05$ for $\pt>200\GeV$.
Electrons (muons) must have isolation $\Irel<0.085$ (0.325).

A BDT is used to distinguish leptons coming from hadron decays and misidentified leptons, collectively denoted ``nonprompt'' leptons, from those produced in the decays of \PW, \PZ, or Higgs bosons (``prompt''). The BDT is trained using MC simulated events and exploits the properties of the jet which contains the lepton, as well as properties of the lepton itself, irrespective of the isolation requirement described above.
It is specifically calibrated for this run, following the same methods as in Refs.~\cite{muonMVA,CMS:2020mpn}: the chosen working point for the BDT score yields more than 99\% prompt lepton efficiency and a nonprompt rejection factor higher than 55\%.
Leptons clustered inside a jet satisfying the loose working point of the \DeepCSV \PQb tagging algorithm~\cite{Sirunyan:2017ezt} are rejected to further suppress nonprompt leptons originating from \PQb hadron decays.
Using this working point with \ttbar events, we obtain a \PQb tagging efficiency of 84\%, and mistag rates of 41\% and 11\% for \PQc and light jets with $\pt>20\GeV$.

\section{Data and simulated samples}
\label{sec:dataandMC}

The measurement is performed using \pp collision data collected during a low-energy (5.02\TeV) and low-intensity LHC run.
The maximum instantaneous luminosity delivered by the LHC during this period was $\mathcal{L}=1.37\times10^{33}\percms$,
and the mean of the pileup distribution was 2, to be compared with the standard conditions in the 13\TeV run in 2017, where $\mathcal{L}=2.07\times10^{34}\percms$, and the mean pileup was 32.

This analysis considers the $t$-channel process as signal, and the following processes as sources of background: single top quark production in the $s$ channel, and in association with a \PW boson (\tW); \ttbar production; quark-antiquark annihilation into lepton-antilepton pairs through \PZ boson or a virtual photon exchange with additional jets (DY+jets); \PW boson production with additional jets (\wjets) and SM events composed uniquely of jets produced through the strong interaction (QCD multijet).
Signal and background contributions are simulated using several MC event generators, but for the QCD multijet background. This contribution is estimated using both MC event generators and control data samples, as further explained in Section~\ref{sec:Background}. The propagation of the generated particles through the CMS detector and the modelling of the detector response in simulations are performed with the \GEANTfour~\cite{Agostinelli:2002hh} package. The pileup distribution in the simulated samples roughly matches that observed in data.

The simulated $t$- and $s$-channel events are generated at next-to-leading order (NLO) in perturbative QCD (pQCD) using \POWHEG~(v2)~\cite{Frixione:2007vw,Alioli:2010xd,Frixione:2007nw,Nason:2004rx}.
A nominal top quark mass (\mtop) of 172.5\GeV and the five-flavour scheme is used. The NNPDF3.1~\cite{Ball:2017nwa} NNLO set is used as the default PDF set. A similar setup is used for the simulation of \tW and \ttbar production. In the case of \tW, the diagram removal scheme~\cite{tWDR} is employed to treat the interference with \ttbar events, using a fixed scale scheme.

The \MGvATNLO (v2.4.2) generator~\cite{Alwall:2014hca} is used to simulate DY+jets and \wjets processes. The simulation is performed at NLO in pQCD and includes up to two extra partons at the matrix element (ME) level. The FxFx matching scheme~\cite{Frederix:2012ps} is used to match jets from the ME calculations to the parton shower (PS). For all processes, simulated events are interfaced with \PYTHIA (v8.230)~\cite{Sjostrand:2014zea},
using the CP5 tune~\cite{Sirunyan:2019dfx} for PS, hadronization, and the
underlying event description.

The individual predictions of the cross sections of single top quark (\tq) and single top antiquark (\tbarq) production in the $t$ channel are $\sto=20.3\,^{+0.5}_{-0.4}\pb$ and $\santitop=10.0\,^{+0.2}_{-0.3}\pb$, respectively.
The cross sections were calculated with the \MCFM program (v10.1)~\cite{Campbell:2020fhf} and next-to-next-to-leading order (NNLO)
in quantum chromodynamics (QCD), using \mtop as 172.5\GeV and the five-flavour scheme, in which the bottom quark is treated as a massless parton constituent of the proton. The quoted uncertainties include the ones due to
the choice of the renormalization (\mur) and the factorization (\muf) scales (their nominal values being \mur= \muf= 175\GeV), and the ones in the proton PDFs, and in the value of the strong coupling constant \asmz, where $m_{\PZ}$ is the mass of the \PZ boson. The combined PDF and \alpS uncertainties were determined at 68\% confidence level (\CL) according to the Hessian
representation of the PDF4LHC21 PDF set~\cite{PDF4LHCWorkingGroup:2022cjn}.

The cross section prediction used for the normalization of \ttbar events is calculated with the \textsc{top++} program~\cite{Czakon:2011xx} at NNLO in pQCD including soft-gluon resummation at next-to-next-to-leading logarithmic (NNLL) accuracy~\cite{Czakon:2013goa}. The SM prediction for \tW events is calculated at approximate NNLO accuracy,
meaning an NLO calculation for QCD order of approximation, matched with soft gluon resummation~\cite{Kidonakis:2021vob} at NNLL, where a logarithmic expansion of soft and collinear gluon radiation is used.
For \wjets and DY+jets, the calculation is performed at NNLO in QCD and NLO in electroweak (EW) orders of approximation~\cite{Melnikov:2006kv}.
Finally, the calculation for $s$-channel events is done at the same order of approximation as for the $t$-channel signal, described in Section~\ref{sec:intro}.
The signal and background simulated samples used in the analysis are summarized in Table~\ref{tab:mc}, as well as their predicted cross sections,
reported in the best available fixed-order calculations and used to normalize the simulated events.

\begin{table}[!ht]
\centering
\topcaption{%
    Summary of simulated samples used to model the signal and background processes. The last column corresponds to the pQCD and EW order of approximation used to normalize the distributions provided by the
    generators.
}
\label{tab:mc}
\renewcommand{\arraystretch}{1.3}
\cmsTable{\begin{tabular}{cccc}
    \multirow{2}{*}{Process} & \multirow{2}{*}{Generator + PS} & Cross section, & \multirow{2}{*}{Order of \snorm approximation} \\[-5pt]
    & & \snorm (pb) & \\
    \hline
    $t$ channel & \POWHEG + \PYTHIA{}8 & $30.3\pm0.6$ & NNLO~\cite{Campbell:2020fhf} \\
    \ttbar & \POWHEG + \PYTHIA{}8 & $69.5\pm3.6$ & NNLO + NNLL~\cite{Czakon:2011xx,Czakon:2013goa} \\
    \tW & \POWHEG + \PYTHIA{}8 & $6.54\pm0.36$ & Approximate NNLO~\cite{Kidonakis:2021vob} \\
    $s$ channel & \POWHEG + \PYTHIA{}8 & $2.74\pm0.07$ & NNLO~\cite{updatedxsecs} \\
    \wjets & \MGvATNLO + \PYTHIA{}8 & $21\,160\pm330$ & NNLO [QCD] + NLO [EW]~\cite{Melnikov:2006kv} \\
    DY+jets & \MGvATNLO + \PYTHIA{}8 & $3\,647\pm63$ & NNLO [QCD] + NLO [EW]~\cite{Melnikov:2006kv} \\
\end{tabular}}
\end{table}

\section{Event selection and background discrimination}
\label{sec:EventSelection}

The final state targeted by this analysis consists of a lepton (electron or muon) and a neutrino from the \PW boson decay, a jet from the \PQb quark hadronization, and an additional forward jet from the hadronization of a light quark.
Candidate events are, consequently, required to contain exactly one selected electron or muon, and at least two jets. Only jets with $\pt>25\GeV$, $\abseta<4.7$, and whose \DR distance from any lepton is greater than 0.4 are considered. In order to reduce the background contamination from QCD multijet events, the magnitude of \ptvecmiss must be greater than 30\GeV, and the transverse mass of the lepton and missing transverse momentum (\mtw) must be greater than 50\GeV. It is defined as:
\begin{equation}
    \mtw=\sqrt{2\pt^{\Pell}\ptmiss(1-\cos\Delta\phi)},
\end{equation}
where $\pt^{\Pell}$ is the \pt of the selected lepton and $\Delta\phi$ is the azimuthal angle between the lepton and \ptvecmiss. To further increase the purity of the signal, events are required to have $\httprime>170\GeV$, where \httprime is defined as the scalar sum of \ptmiss and the \pt of the jets and of the lepton. The region defined by these criteria is known as the signal region (SR).
Events are classified according to their jet and \PQb-tagged jet multiplicities: those considered for the measurement have exactly two reconstructed jets, only one of which is \PQb tagged, or exactly three reconstructed jets,
with either one or two of them that are \PQb tagged. These categories are denoted as \njnb21, \njnb31, and \njnb32, respectively. The two latter categories are intended to improve constraints on the \ttbar background. All the categories are futher split according to the flavour and the electric charge of the selected
lepton in the event, resulting in twelve mutually exclusive analysis categories.

\begin{figure}[!htp]
\centering
\includegraphics[width=0.55\textwidth]{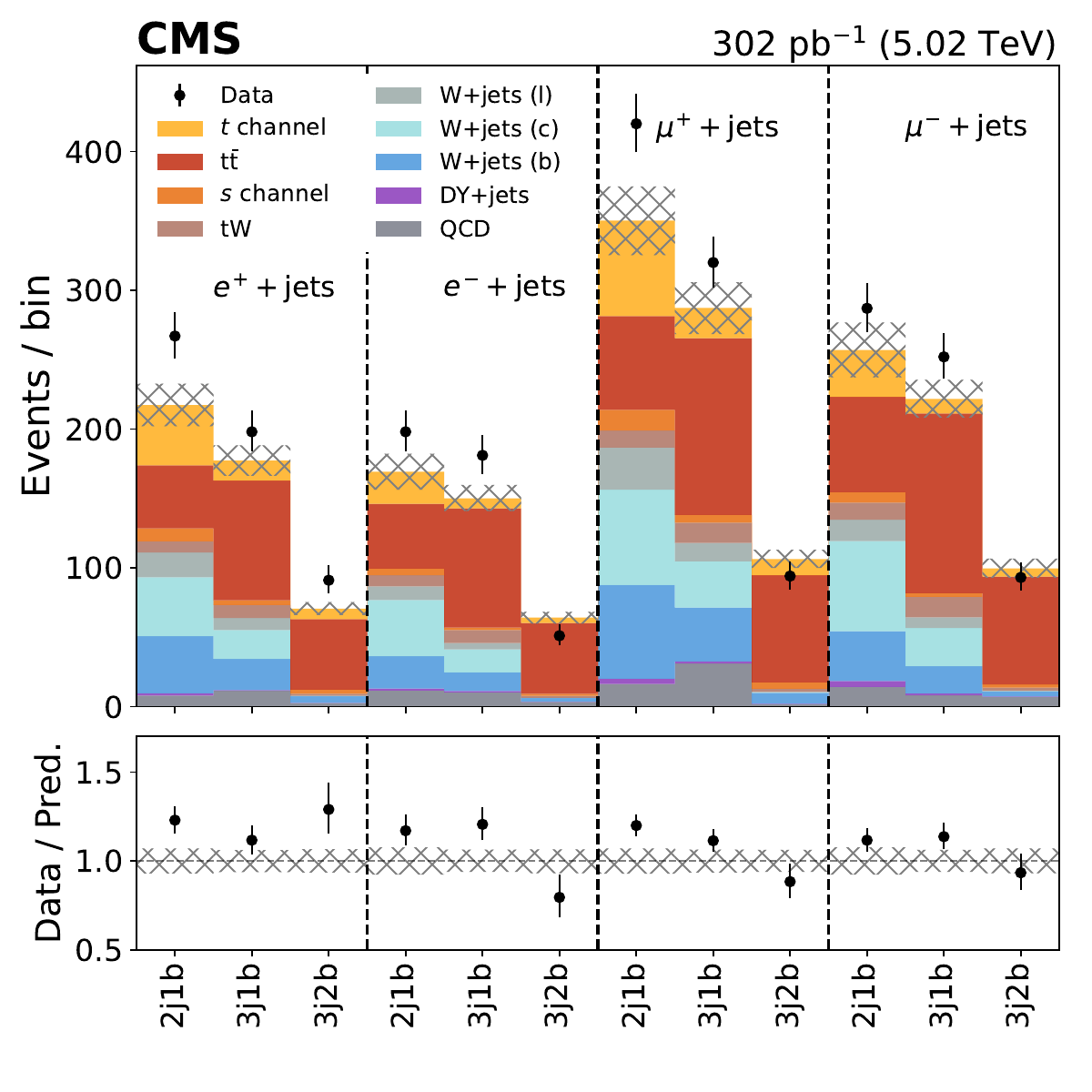}
\caption{Observed and predicted number of events in each of the twelve categories considered in the analysis, before the ML fit. The vertical error bars represent the statistical uncertainty associated to the data, and the hatched band the  uncertainty in the prediction. All uncertainties considered in the analysis are included in the uncertainty band. The lower panels show the data-to-prediction ratio. The normalizations of the processes are those of the SM predictions except for QCD, which is estimated from data. The \wjets background is split as explained in Section~\ref{sec:Background}.}
\label{fig:master}
\end{figure}

\begin{figure}[!htp]
\centering
\includegraphics[width=0.46\textwidth]{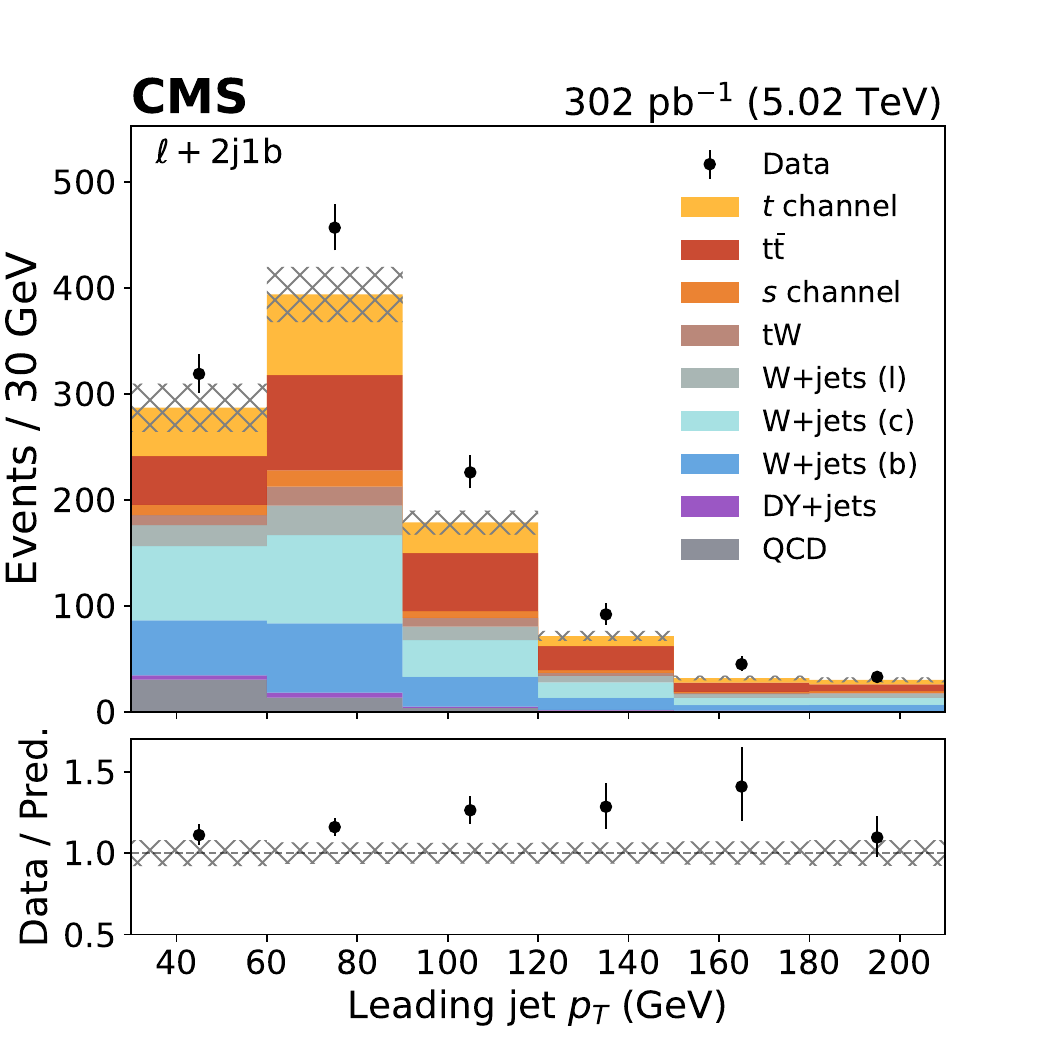}
\hfill
\includegraphics[width=0.46\textwidth]{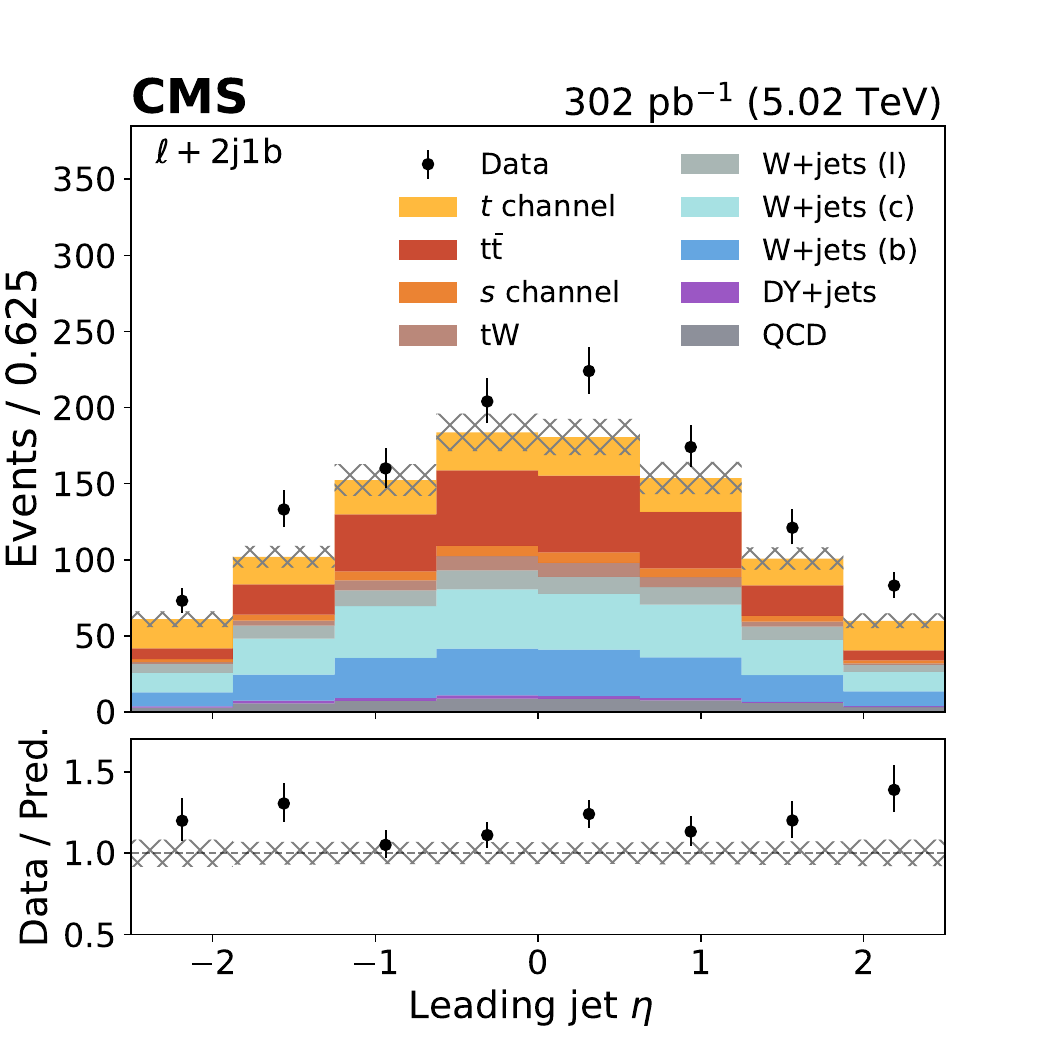}
\caption{Observed and predicted distributions of the leading jet \pt (left) and $\eta$ (right), in the \njnb21 category, before the ML fit. The vertical error bars represent the statistical uncertainty associated to the data, and the hatched band the uncertainty in the prediction. The lower panels show the data-to-prediction ratio. The first and last bins in each distribution include underflow and overflow events, respectively. The normalizations of the processes are those of the SM predictions except for QCD, which is estimated from data. The \wjets background is split as explained in Section~\ref{sec:Background}.}
\label{fig:control}
\end{figure}

The resulting observed number of events in data and the expected number of events for signal and background in each
of the twelve categories of the analysis are shown in Fig.~\ref{fig:master}. Figure~\ref{fig:control} shows the leading jet (in \pt) transverse momentum and pseudorapidity inclusively in lepton charge and flavour for the \njnb21 jet category, which contains the majority of the signal events.

\subsection{Top quark reconstruction}
\label{sec:topreco}

Top quarks (antiquarks) give rise to positively (negatively) charged leptons. The four-momen\-tum of a top quark or antiquark is reconstructed from its decay products: the charged lepton, the \PQb jet, and the unobserved neutrino (\PGn); following the procedure that has already been used in previous CMS analyses~\cite{Sirunyan:2016cdg,top201711}. The reconstruction is done in two steps: first a \PW boson is reconstructed and then it is associated with a \PQb-tagged jet.

To reconstruct the \PW boson, we first reconstruct the unobserved neutrino. We assign the $x$ and $y$ components of the \ptvecmiss to the momentum of the neutrino, and infer its $z$ component by applying the following constraint on the \PW boson mass, \mW:
\begin{equation}\label{eq:constraint}
    \mW^2 = \left(E^{\Pell}+\sqrt{(\ptmiss)^2+(p^{\PGn}_z)^2}\right)^2 - \left(p_x^{\Pell}+p^{\text{miss}}_{\mathrm{T},x}\right)^2 - \left(p_y^{\Pell}+p^{\text{miss}}_{\mathrm{T},y}\right)^2 - \left(p_z^{\Pell}+p_z^{\PGn}\right)^2.
\end{equation}
This equation leads to the following two solutions:
\begin{equation}\label{eq:neutrino}
    p_z^{\PGn} = \frac{\Lambda p_z^{\Pell}}{\big(\pt^{\Pell}\big)^2} \pm \sqrt{\frac{\big(\Lambda p_z^{\Pell}\big)^2}{\big(\pt^{\Pell}\big)^4} - \frac{\big(E^{\Pell}\ptmiss\big)^2-\Lambda^2}{\big(\pt^{\Pell}\big)^2}},
\end{equation}
where
\begin{equation}
    \Lambda = \frac{\mW^2}{2} + p_x^{\Pell}p^{\text{miss}}_{\mathrm{T},x} + p_y^{\Pell}p^{\text{miss}}_{\mathrm{T},y}.
\end{equation}
There can either be two real solutions for $p_z^{\PGn}$, or the solutions have an imaginary component. In the first case, the solution with the smallest absolute value is chosen. In the second, an extra constraint is adopted: the imaginary component is eliminated by replacing $p^{\text{miss}}_{\mathrm{T},x}$ and $p^{\text{miss}}_{\mathrm{T},y}$ with modified values $p_x^{\PGn}$ and $p_y^{\PGn}$ such that $\mtw=\mW$, but still respecting the constraint from Eq.~\eqref{eq:constraint}. This is achieved by setting the discriminant in Eq.~\eqref{eq:neutrino} to zero. This leads to a quadratic relation between $p_x^{\PGn}$ and $p_y^{\PGn}$ with two possible solutions: the one with the smallest difference between $\pt^{\PGn}$ and \ptmiss is chosen.

The \PW boson is reconstructed from the neutrino candidate and the selected lepton. In the \njnb21 and \njnb31 regions, the \PQb-tagged jet is associated with the \PQb quark from the top quark decay and the non-\PQb-tagged jet with the highest \pt is associated with the recoiling light quark. In the \njnb32 region, the \PQb-tagged jet with the larger \PQb tag discriminator value is associated with the \PQb quark from the top quark decay. Finally, the top quark is reconstructed from the \PW boson candidate and the selected \PQb-tagged jet. Its mass is then used as an input variable to the multivariate analysis (MVA) described in Section~\ref{sec:mvadiscriminator}.

\subsection{Multivariate analysis discriminator}
\label{sec:mvadiscriminator}

\begin{figure}[!ht]
\centering
\includegraphics[width=0.46\textwidth]{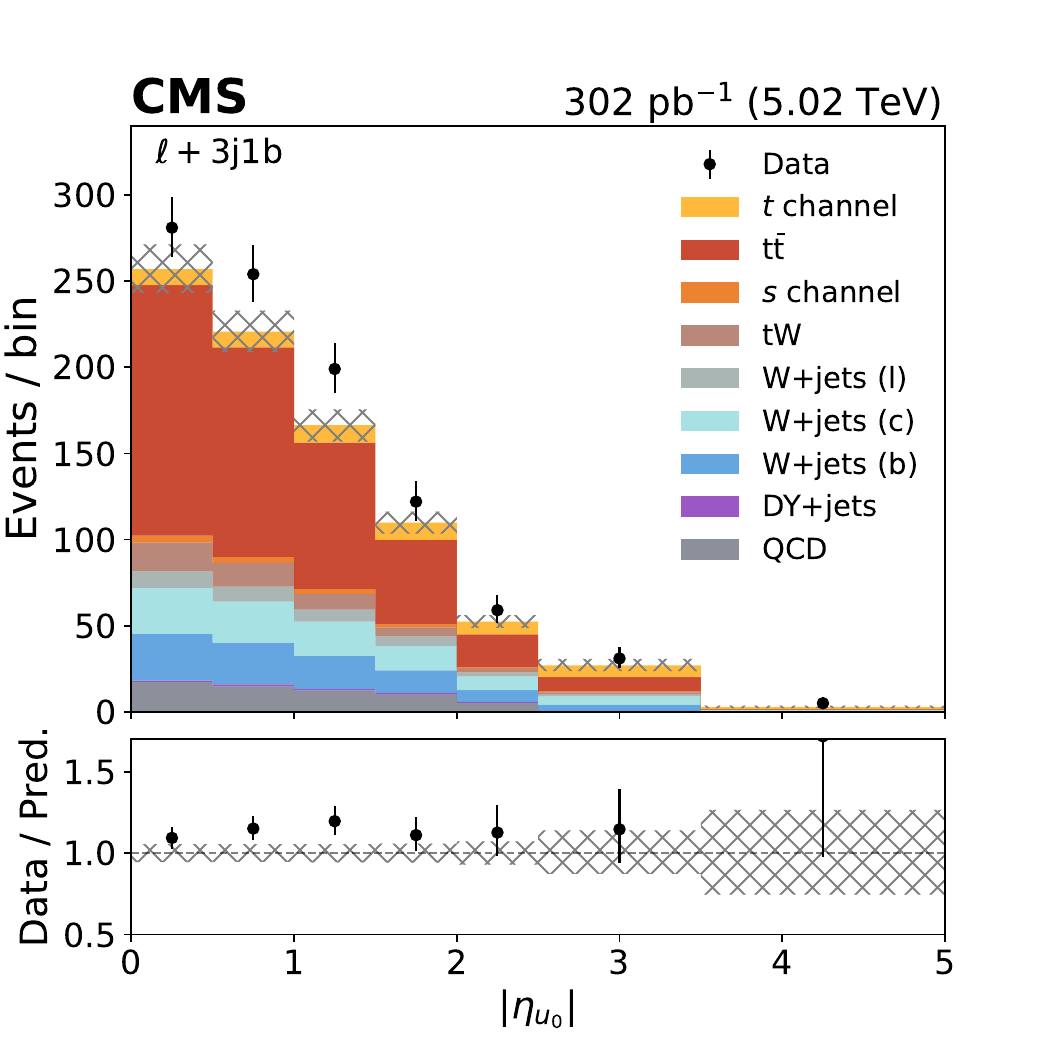}
\hfill
\includegraphics[width=0.46\textwidth]{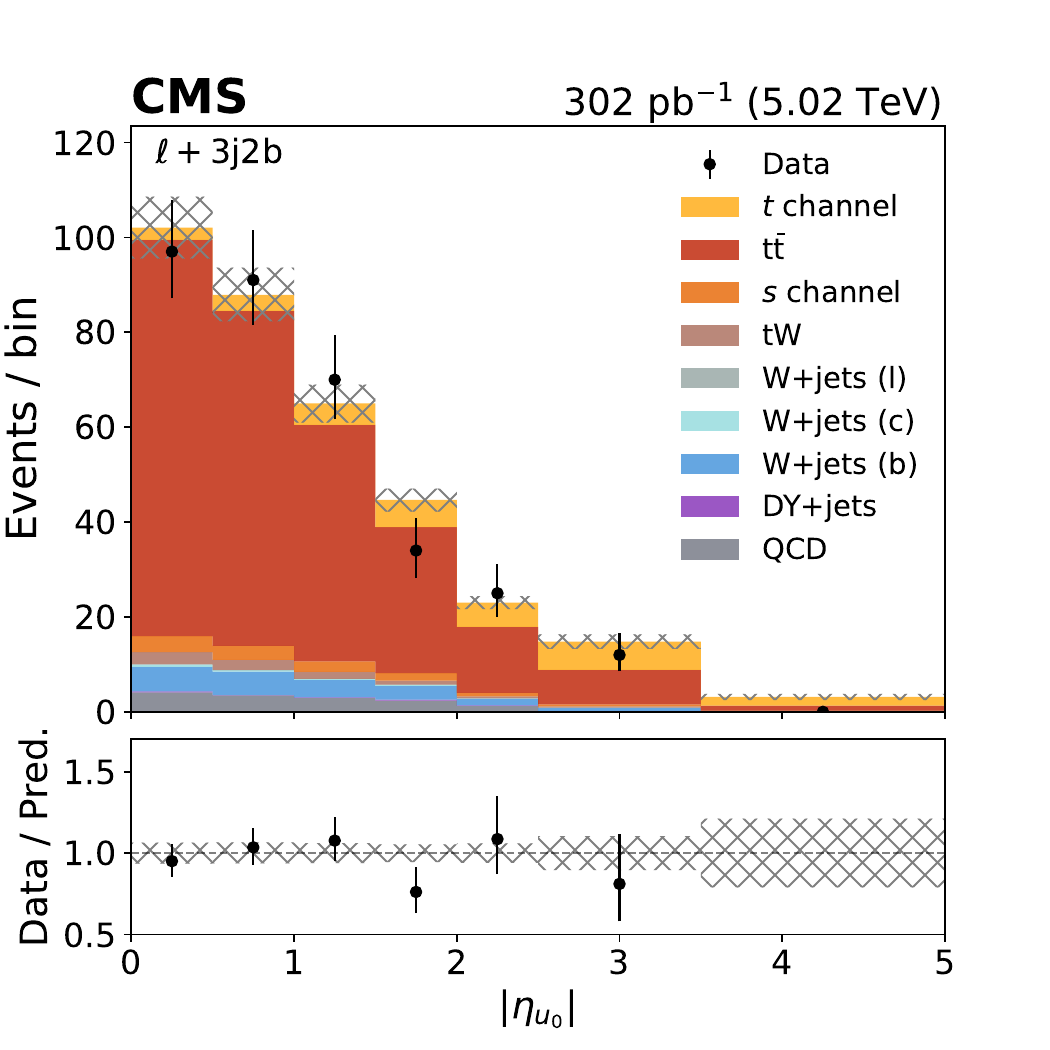}
\caption{Observed and predicted distributions of \absueta in the \njnb31 (left) and \njnb32 (right) categories, before the ML fit. The vertical error bars represent the statistical uncertainty associated to the data, and the hatched band the uncertainty in the prediction. The lower panels show the data-to-prediction ratio. The normalizations of the processes are those of the SM predictions except for QCD, which is estimated from data. The \wjets background is split as explained in Section~\ref{sec:Background}.}
\label{fig:absu0eta3j}
\end{figure}

In the \njnb21 categories, which contain most of the signal events, \wjets events constitute the largest source of background, while \ttbar dominates in the other categories. A dedicated MVA
binary classifier has been developed for these categories in order to better discriminate signal from these backgrounds, while for the other categories a single observable, the absolute value of $\eta$ of the untagged jet (\absueta), is used. Figure~\ref{fig:absu0eta3j} shows the distribution of the latter variable in those categories.

As for the MVA, a random forest~\cite{cutler2012random} has been trained inclusively in lepton charge and flavour, using a mix of \ttbar and \wjets events as background sample and $t$-channel events as signal.
Subsets of the simulated samples, orthogonal to those used in the rest of the analysis have been employed for this purpose: 70\% of these events are used for the training and the remaining ones for the testing. The variables used in the classifier are described in Table~\ref{tab:variablesmva}, ordered by their discriminating power.

\begin{table}[!ht]
\centering
\topcaption{%
    Summary of the variables used for training the MVA classifier in the \njnb21 categories. The variables are ordered according to their discriminating power, assessed via the mean decrease in impurity method~\cite{cutler2012random}. ``Leading'' jet refers to the jet with the highest \pt in the event.
}
\label{tab:variablesmva}
\renewcommand{\arraystretch}{1.15}
\begin{tabular}{cp{0.77\textwidth}}
    Variable & Definition \\
    \hline
    \absueta & The absolute value of $\eta$ of the untagged jet \\
    \deltaeta & The absolute value of the difference in $\eta$ between the
    tagged and untagged jets \\
    \mtw & The transverse mass of \Pell and \ptmiss \\
    \mtopreco & The invariant mass of the reconstructed top quark \\
    $\Delta R(\mathrm{j,j'})$ & The $\Delta R$ between the two jets \\
    \mjj & The invariant mass of the two jets \\
    \mlb & The invariant mass of the lepton and the \PQb-tagged jet \\
    \sumvecall & The scalar sum of the \pt of all jets and the lepton in the
    event \\
    \httprime & The scalar sum of \ptmiss and the \pt of the jets and of the
    lepton in the event \\
    \detalu & The absolute value of the difference in $\eta$ between the
    lepton and the untagged jet \\
    \dphiwub & The absolute value of the difference in $\phi$ between the
    reconstructed \PW~boson and the jet pair \\
    \dptwub & The absolute value of the difference in \pt between the
    reconstructed \PW~boson and the jet pair \\
\end{tabular}
\end{table}

\begin{figure}[!ht]
\centering
\includegraphics[width=0.46\textwidth]{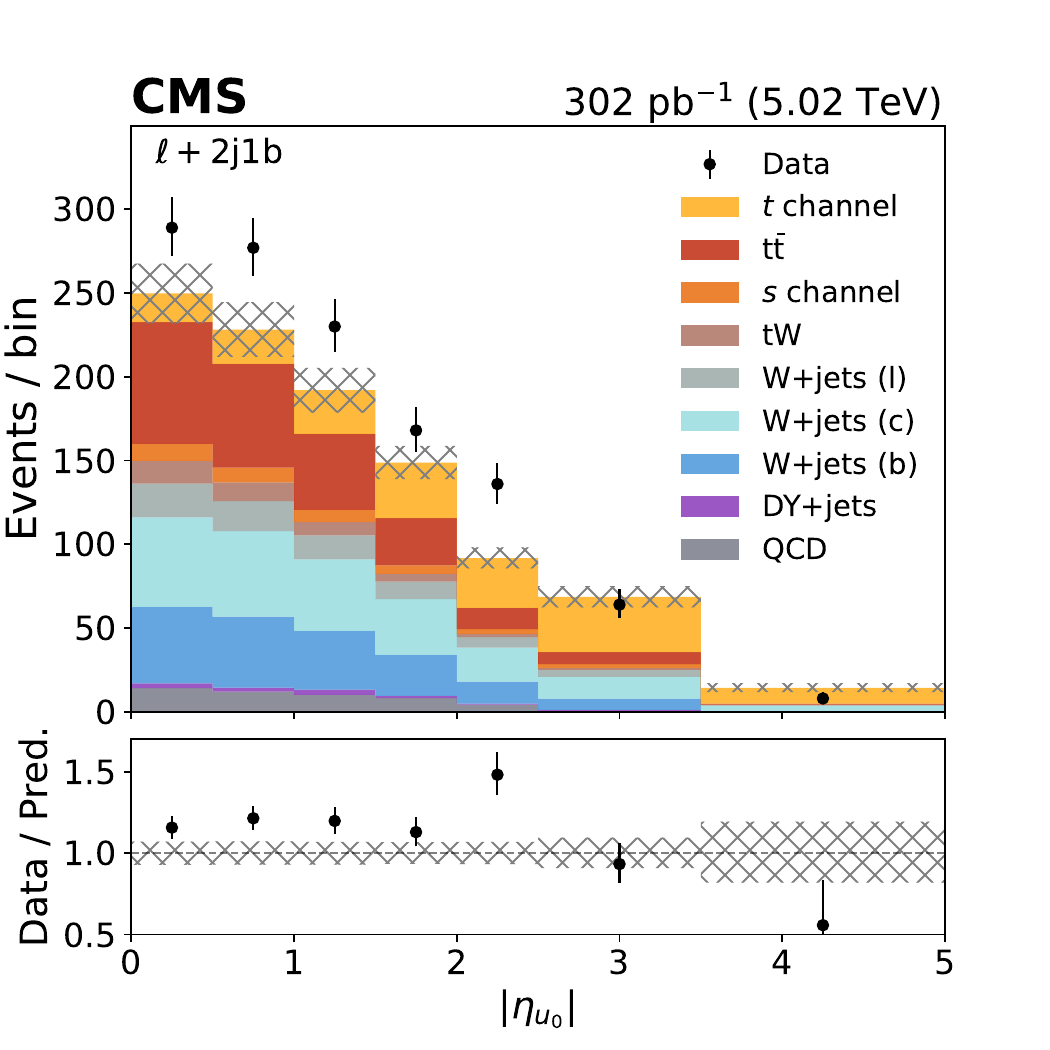}
\hfill
\includegraphics[width=0.46\textwidth]{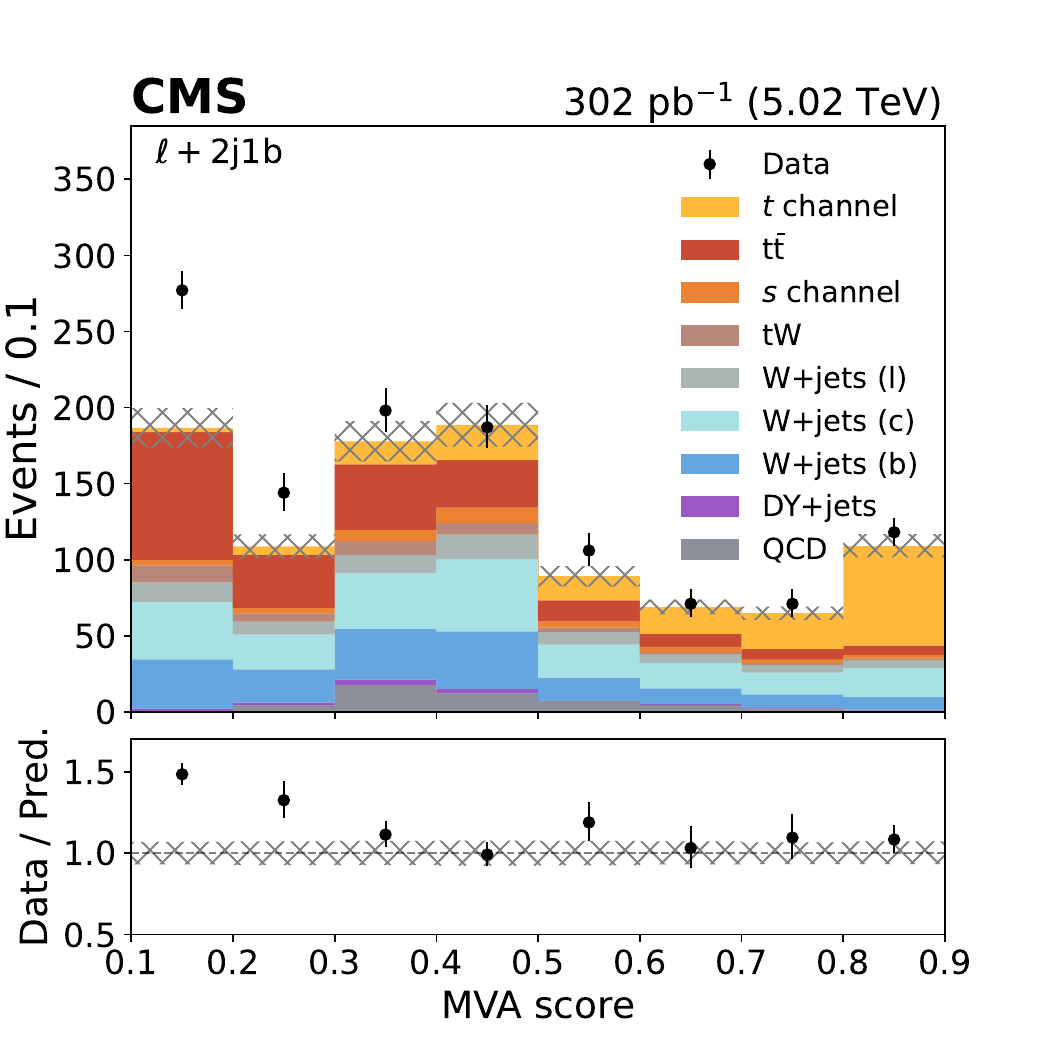}
\caption{Observed and predicted distributions of the most discriminating input variable used in the random forest training, \absueta (left), and the output MVA discriminator (right), in the \njnb21 category, before the ML fit. The vertical error bars represent the statistical uncertainty associated to the data, and the hatched band the uncertainty in the prediction. The lower panels show the data-to-prediction ratio. The first and last bins in the MVA output distribution include underflow and overflow events, respectively. The normalizations of the processes are those of the SM predictions except for QCD, which is estimated from data. The \wjets background is split as explained in Section~\ref{sec:Background}.}
\label{fig:inputsdatamc}
\end{figure}

Figure~\ref{fig:inputsdatamc} shows the comparison of data and prediction for the most discriminating input variable of the MVA, the absolute value of the $\eta$ of the leading untagged jet (\absueta), and the output MVA discriminator in the \njnb21 category. A reasonable agreement is observed between data and predictions for all variables used in the classifier before the fit.

\section{Background estimation}
\label{sec:Background}

The event selection is designed to maximize the signal contribution over the expected background contamination. The background contributions come from the following processes: \ttbar, \tW, \wjets production in the single-lepton final state, QCD multijet events when a jet is misidentified as a lepton, and DY+jets events. 
The estimation of these background contributions is based on MC simulation, except for the QCD multijet contribution, which is determined in a hybrid approach, exploiting control samples in data as described later.
The \wjets background is split according to the generator level flavour of the additional jets into \wb jets if at least one \PQb jet is found in the analyzed event, \wc jets if there are no \PQb jets but there is at least one \PQc jet, and \wlight jets (including light quark or gluon jets) otherwise. The split is included to handle better flavour dependent tagging efficiencies and production uncertainties. 
As a further check of this process, a validation region has been defined by inverting the $\httprime>170\GeV$ requirement, thus enhancing the contribution of \wjets events. In this region, good agreement between data and prediction is found.

In order to estimate the QCD multijet contribution, a two-step procedure is applied, using a control region (CR) defined following the same criteria as the SR, but with the lepton failing either the \Irel or prompt lepton BDT requirements described in Section~\ref{sec:CMS}.
In the first step, the number of QCD background events in the SR (\nSR{QCD}) is estimated for each of the twelve event categories. To do so, extrapolation factors from the CR to SR are obtained from a sample of events with
$\ptmiss<25\GeV$, where a large contribution of QCD background and low signal contamination is observed. The values for \nSR{QCD} are calculated as
\begin{equation}\label{eq:qcdextrapolation}
    \nSR{QCD} = \Big(\nCR{obs}-\nCR{MC}\Big)_{\text{yields}}
    \frac{\nSR{obs, low \ptmiss}-\nSR{MC, low \ptmiss}}{\nCR{obs, low \ptmiss}-\nCR{MC, low \ptmiss}},
\end{equation}
where the total normalization is expressed as the product of two components: an extrapolation factor (second term on the right-hand side), which accounts for the transfer from the CR to SR, and a category-specific normalization factor, estimated from events in the CR that satisfy the nominal \ptmiss requirement (first term on the right-hand side). Both factors are determined after subtracting the contributions, derived from the simulated samples, of the signal and the other background sources considered (\nSR{MC}), from the observed numbers of events in data (\nSR{obs}).

In a second step, the shapes of variable distributions are determined from QCD multijet simulated samples. For each distribution, a template used in all categories is obtained by summing all the contributions to all the regions. Here, the selection requirements in \ptmiss, \mtw, and \httprime described in Section~\ref{sec:EventSelection} are removed
in order to reduce the statistical uncertainty on the prediction having checked that no significant bias is introduced. Individual distributions (per category) are then obtained by normalizing these shapes to the yields obtained in the previous step. The whole procedure is an improved alternative of the one followed in other analysis using this data set~\cite{CMS:2024ghc}.

To estimate the normalization uncertainty in the QCD multijet background, the normalization factor is re-evaluated after shifting the \ptmiss requirement by $\pm$5\GeV in both the SR and CR, and calculating the difference in acceptance, similarly to other analysis using this data set~\cite{CMS:2024ghc}. This leads to a flat normalization uncertainty of 30 (40)\% in the \mjets (\ejets) channel, covering the observed variation. A shape uncertainty is assessed by recalculating the QCD contribution using alternative extrapolation factors, determined independently in bins of the leading lepton \pt and \abseta. The QCD multijet background accounts for approximately 7\% of the total event yield.

\section{Systematic uncertainties}
\label{sec:Systematics}

The measurement of \sboth is affected by sources of systematic uncertainty related to the detector effects, background estimation, and theoretical assumptions.
The statistical uncertainty coming from the limited size of the simulated samples is incorporated into the ML fit using the Barlow--Beeston ``light'' method~\cite{Conway:2011in}.
The experimental uncertainties (considered for all processes) and modelling uncertainties (considered only for the $t$-channel signal and \ttbar background processes) affect both the acceptance and efficiency. The modelling uncertainties account for the choice of \mur and \muf scales, PS scale, ME and PS matching, PDFs, and underlying event tune. Furthermore, a normalization uncertainty is considered for each background process. The total uncertainty is calculated by combining the effects of all the individual systematic components, assuming they are independent. The sources of systematic uncertainties and the methods for their estimation are outlined below.
\begin{description}
\item[Lepton-related uncertainties:]
Lepton reconstruction and identification efficiencies are measured using DY events in simulation and low-pileup \pp collision data at 5.02\TeV, similarly to what is described in Refs.~\cite{CMS:2020uim}. Corrections are derived using the ``tag-and-probe'' method~\cite{CMS:2010svw} as functions of the lepton \pt and $\eta$, and are applied to the simulation to match the efficiencies in data. The uncertainty in these corrections is estimated through the variation of different parameters in the fitting templates of the method, as well as in the tag selection criteria, and is propagated to the measurements as one single source per lepton flavour, including the statistical and systematic uncertainties. It is around 0.5\% for both lepton flavours.

The HLT trigger efficiency is estimated using a cross-trigger technique. This consists in measuring it in data using the muon trigger as reference for the electron trigger efficiency and vice-versa in the case of muons. The related uncertainty is estimated by combining a statistical contribution, evaluated with the Clopper--Pearson method~\cite{clopper} and a systematic contribution, from simulation to data efficiency ratios in several lepton \pt and \abseta bins. It is found to be around 1--3\% for electrons and 0.8--1.6\% for muons.

\item[Jet-related uncertainties:] The contribution of the uncertainty in the JES is parametrized as a function of jet \pt and $\eta$, usually resulting in variations of no more than a few percent as described in Ref.~\cite{Khachatryan:2016kdb}. Fourteen different sources of uncertainties are considered, and treated as independent nuisance parameters.
The uncertainty due to the accuracy of the jet energy resolution (JER) is determined by changing the JER correction scale factor from its nominal value by $\pm$1 standard deviation (s.d.), depending also on the \pt and $\eta$ of the jet.

\item[\ptmiss-related uncertainty:] The uncertainty in \ptmiss from the contribution of unclustered energy (energy which is not aggregated into a specific cluster in the calorimeters) is evaluated based on the momentum resolution of the different PF candidates. In addition, JER, JES, and lepton energy scale effects are propagated to \ptmiss.

\item[\PQb tagging uncertainties:] In order to match the efficiency of the \PQb tagging algorithm evaluated on simulated samples to that observed in data, appropriate scale factors must be applied as a function of \pt and \abseta of the jets. These scale factors are derived with an associated uncertainty, coming from different sources such as JES and JER. Their effect in the analysis is estimated by varying the scale factors by their uncertainties, and propagating the effect to all the categories in the analysis. Tagging and mistagging are taken as two independent sources of uncertainty.

\item[L1 trigger efficiency:] During the data taking, a gradual shift in the timing of the inputs of the ECAL L1 trigger in the region $\abseta>2$ caused a trigger inefficiency~\cite{Sirunyan:2020zal}. Simulations are corrected to mimic this behaviour observed in data. This correction has an inherent uncertainty associated to the method used to calculate it, which is propagated to the measurements by varying the correction by its associated uncertainty.

\item[Scale choice:] The uncertainty related to the missing higher-order diagrams in \POWHEG is estimated by independently varying each of the \muf and \mur choices a factor 2 or 1/2, excluding combinations with variations in opposite directions.

\item[PS scale:] The effect of the choice of PS scale is studied by changing separately the scale used for the initial- and final-state radiation by a factor 2 and 1/2 with respect to its nominal value.

\item[ME and PS matching:] The effect of the ME and PS matching is taken into account using the \hdamp~parameter of the \POWHEG generator, with a nominal value of $(1.4\,^{+0.9}_{-0.5})\,\mtop$~\cite{Sirunyan:2019dfx}. The uncertainty is estimated by varying this parameter, using dedicated signal samples. The variation with respect to the central value of the signal acceptance is taken as the uncertainty. Due to the limited size of the dedicated simulated samples, a smoothing method following the lowess technique is applied~\cite{lowess}.

\item[Parton distribution functions:] The uncertainty arising from the choice of the proton PDFs is determined by reweighing the sample of simulated signal events according to 100 replicas from the NNPDF3.1 NNLO PDF set~\cite{Butterworth:2015oua}. An extra contribution from the variation of \asmz corresponding to its uncertainty is considered.

\item[Underlying event tune:] The parameters of \PYTHIA related to the underlying event tune~\cite{Sirunyan:2019dfx} are adjusted to match observations in data. We consider changes in these parameters as sources of uncertainties and we use dedicated signal samples accounting for these changes. The variation with respect to the central value of the signal acceptance is taken as the uncertainty. Due to the limited size of the dedicated samples, a smoothing method following the lowess technique is applied~\cite{lowess}.

\item[Background normalization:] The uncertainty in the \tW production cross section is taken to be 5.6\%, and 5.0\% is used both for the \ttbar production, and the single top quark production in the $s$ channel, based on the theoretical uncertainties~\cite{Kidonakis:2021vob,Czakon:2013goa, updatedxsecs}. A normalization uncertainty of 40 (30)\% is applied to the QCD background in the \ejets (\mjets) categories and decorrelated per lepton flavour, and jet and \PQb-tagged jet multiplicity bin. Normalization uncertainties of 20\% are assigned in the \wjets background estimates (for the three \wlight, \wc, and \wb contributions). A 30\% normalization uncertainty is assigned to the DY+jets background estimate. These uncertainties are designated following the prescription of other analyses using the same data set, since the detector conditions and analysis strategy are very similar to the ones presented in this work~\cite{Sirunyan:2017ule,wjetsref}.

\item[Pileup and integrated luminosity:]  Uncertainties related to pileup are not considered since the number of pileup events is small. The uncertainty in the measurement of the integrated luminosity is estimated to be 1.9\%~\cite{bib:LUM-19-001, CMS:LUM-17-003}.

\item[QCD background estimation method:] A systematic contribution is considered to account for uncertainties in the shape estimation of the distributions for the QCD multijet background, as discussed in Section~\ref{sec:Background}. The uncertainty is decorrelated per lepton flavour, and jet and \PQb-tagged jet multiplicity bin.
\end{description}
Most of these uncertainties, especially those coming from experimental effects, affect both the shape of the distributions used in the fit and their normalization through acceptance. Modelling uncertainties that are calculated with several weights, such as PDF and scale uncertainties, are estimated for each region of the analysis and applied as normalization uncertainties in each of the regions. Uncertainties of the background normalization and of the integrated luminosity uncertainty are treated as global normalization uncertainties.

\section{Signal extraction and results}
\label{sec:Results}

\begin{figure}[!p]
\centering
\includegraphics[width=0.8\textwidth]{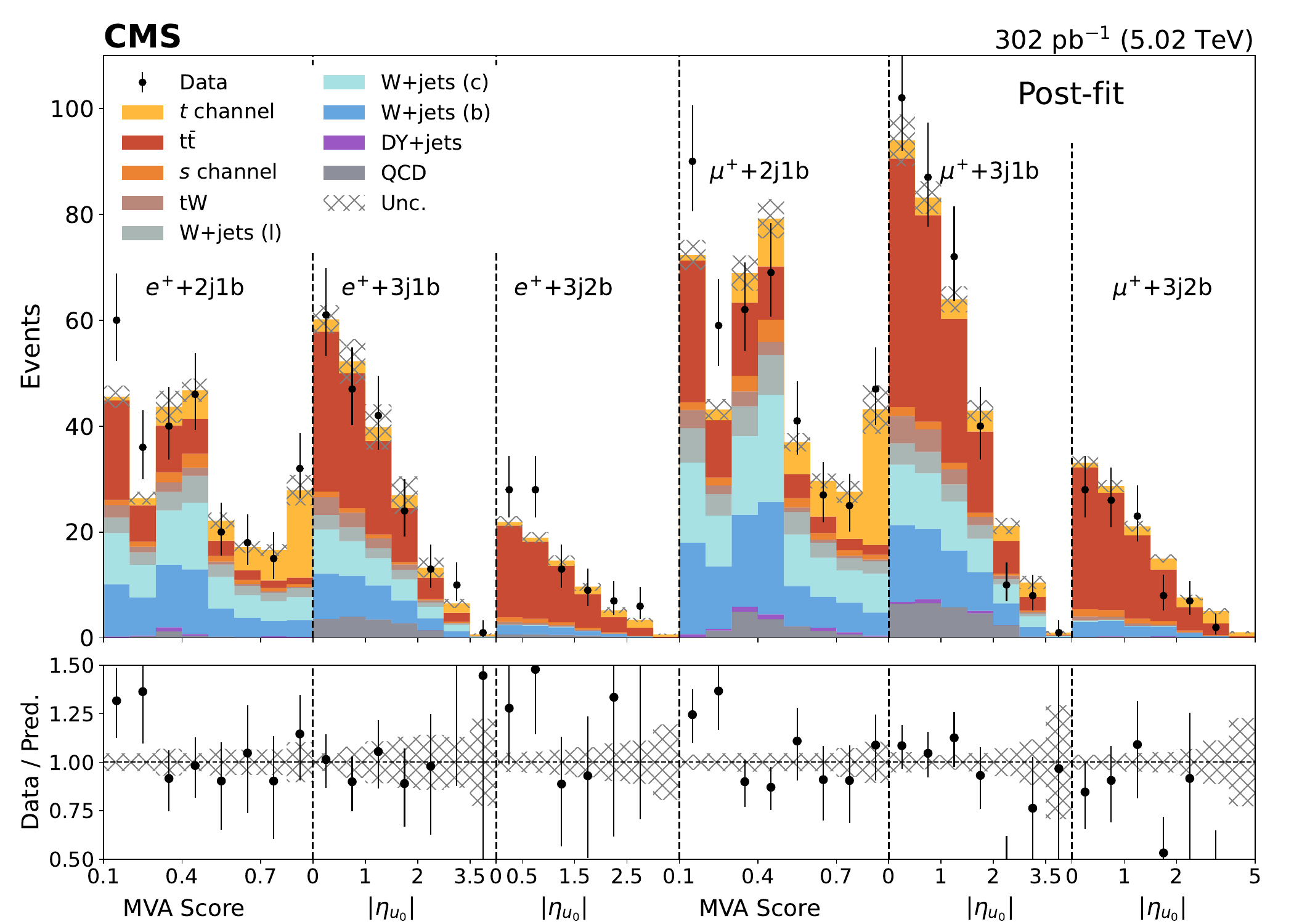}\\
\includegraphics[width=0.8\textwidth]{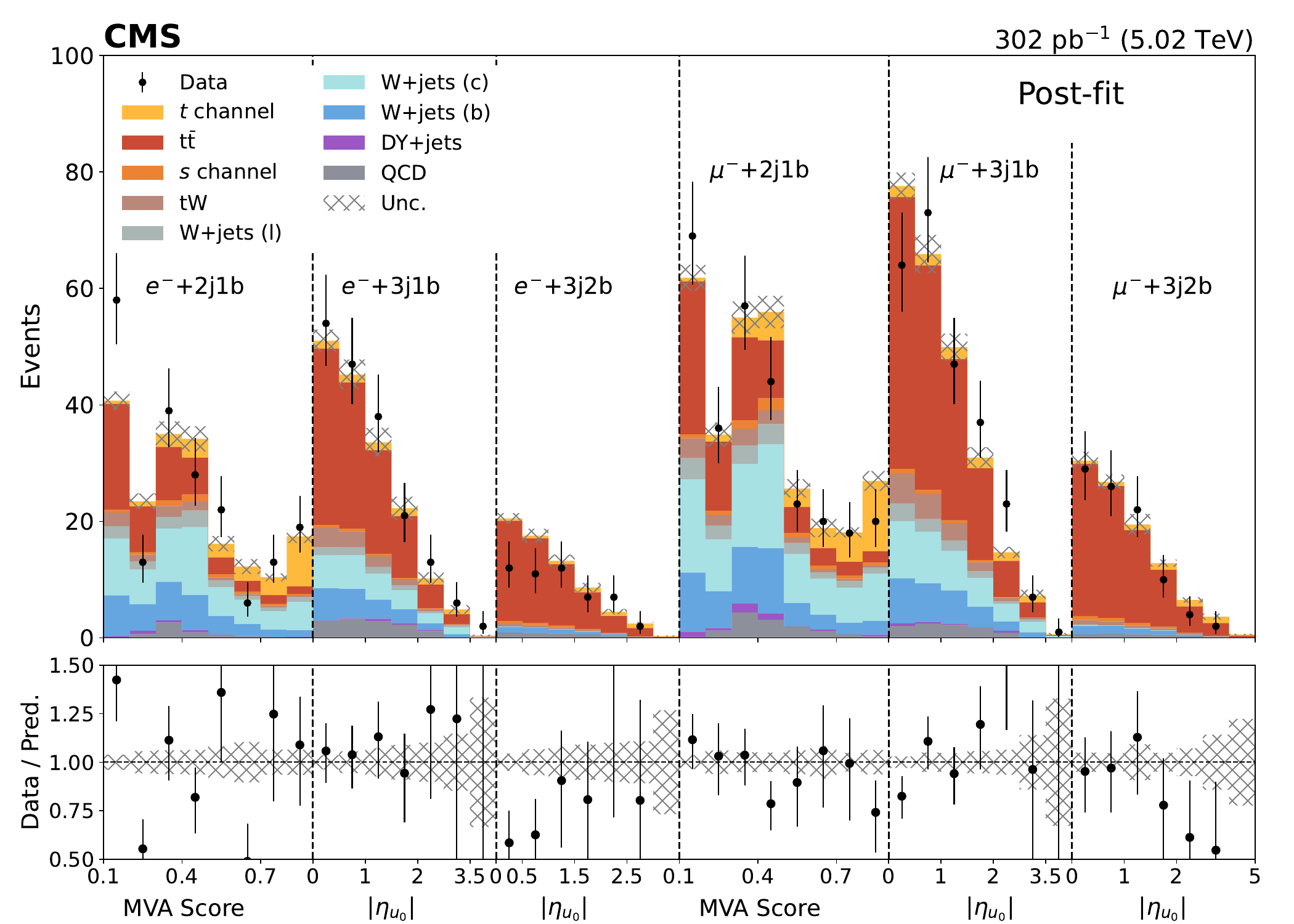}
\caption{Distributions for the $\Pell^+$+jets (upper) and $\Pell^-$+jets (lower) categories where the prediction parameters are set to the values obtained in the fit. They show the MVA score and \absueta distributions, which are the fitted variables in the ML fit in the \njnb21 category and the categories with three jets, respectively. The vertical error bars represent the statistical uncertainty associated to the data, and the hatched band the uncertainty of the prediction, after the fit. All uncertainties considered in the analysis are included in the uncertainty band. The lower panels show the ratio of data to prediction. The first and last bins in the MVA output distributions include underflow and overflow events, respectively.}
\label{fig:fit_lp}
\end{figure}

The cross sections are obtained by performing two independent profile ML fits in the twelve analysis categories, using the CMS \textsc{combine} tool~\cite{cmscollaboration2024cms}. The fits are performed on the distributions of the MVA score for the \njnb21 categories and of \absueta for the others.
The selection of \absueta is motivated by the distinctive feature of the signal process, where the recoiling light quark tends to be produced in the forward region of the detector.
Using the MVA score distribution instead of \absueta for the \njnb21 categories results in an absolute reduction of 8\% in the final uncertainty of the result.
The inclusive cross section \sboth is obtained by fitting a common signal strength modifier ($r$), representing the ratio of the observed signal to the theoretical prediction, to both the \tq and \tbarq processes. In the case of the individual top quark and antiquark cross sections, \sto and \santitop, two independent modifiers are fit. Signal strength modifiers are fitted simultaneously with nuisance parameters related to the systematic uncertainties. In the fits, eight bins are used for the MVA score and seven for the leading untagged jet \abseta distributions, resulting in 88 bins. The MVA score and \absueta distributions for all the categories after the fit are shown in Fig.~\ref{fig:fit_lp}.

\begin{figure}[htp!]
\centering
\includegraphics[width=0.75\textwidth]{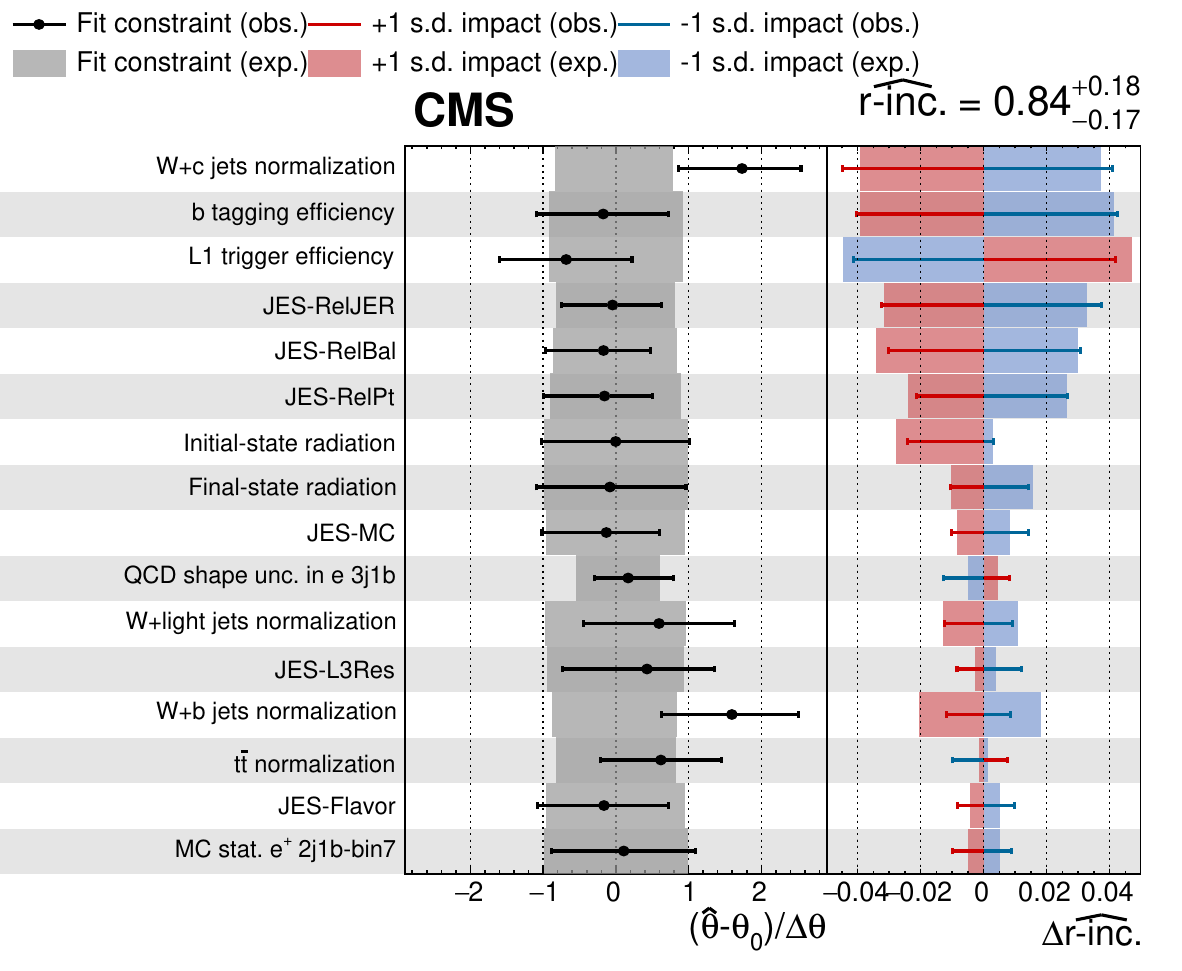}
\caption{The largest impacts of nuisance parameters on the signal strength, $\Delta\widehat{\text{r-inc.}}$ (right column) and ratios $(\hat{\theta}-\theta_0)/\Delta\theta$ (middle column) for the nuisance parameters listed in the left column from the ML fit used to determine the measured \tqandtbarq cross section. The horizontal bars in the middle column show the ratio of the uncertainties of the fit result to the pre-fit ones, indicating the constraint on the nuisance parameter.
The JES uncertainties are divided into several sources related to the various methods used to extract the uncertainties as described in Ref.~\cite{Khachatryan:2016kdb}.}
\label{fig:impactsboth}
\end{figure}

A cross section of $\sboth=25.5\,^{+5.6}_{-5.3}\pb$ is measured, corresponding to a relative uncertainty of around $\pm$21\%, with
a systematic uncertainty of around $\pm$16\%, an integrated luminosity uncertainty of $\pm$1.9\%, and a statistical uncertainty of $\pm$14\%. The dominant sources of uncertainty are those associated
with the \PQb tagging scale factors and the \wc jets normalization. Figure~\ref{fig:impactsboth} shows the impact of the different systematic uncertainties on the signal strength and the constraints on the nuisance parameters. The impact is defined as the shift $\Delta\hat{r}$ induced in $r$ when the nuisance parameter $\theta$ is varied by $\pm$1 s.d.\ around the value obtained after the fit. The figure also shows the ratios $(\hat{\theta}-\theta_0)/\Delta\theta$, where $\hat{\theta}$ and $\theta_0$ are the values of a given nuisance parameter after and before the fit, and $\Delta\theta$ is the corresponding uncertainty before the fit.
In the case of W+c jets normalization, the ratio is displaced from zero because of the tension between data and prediction in the first two bins in some of the MVA score distributions, related to the difficult modelling of the \wjets background.

\begin{figure}[!htp]
\centering
\includegraphics[width=0.48\textwidth]{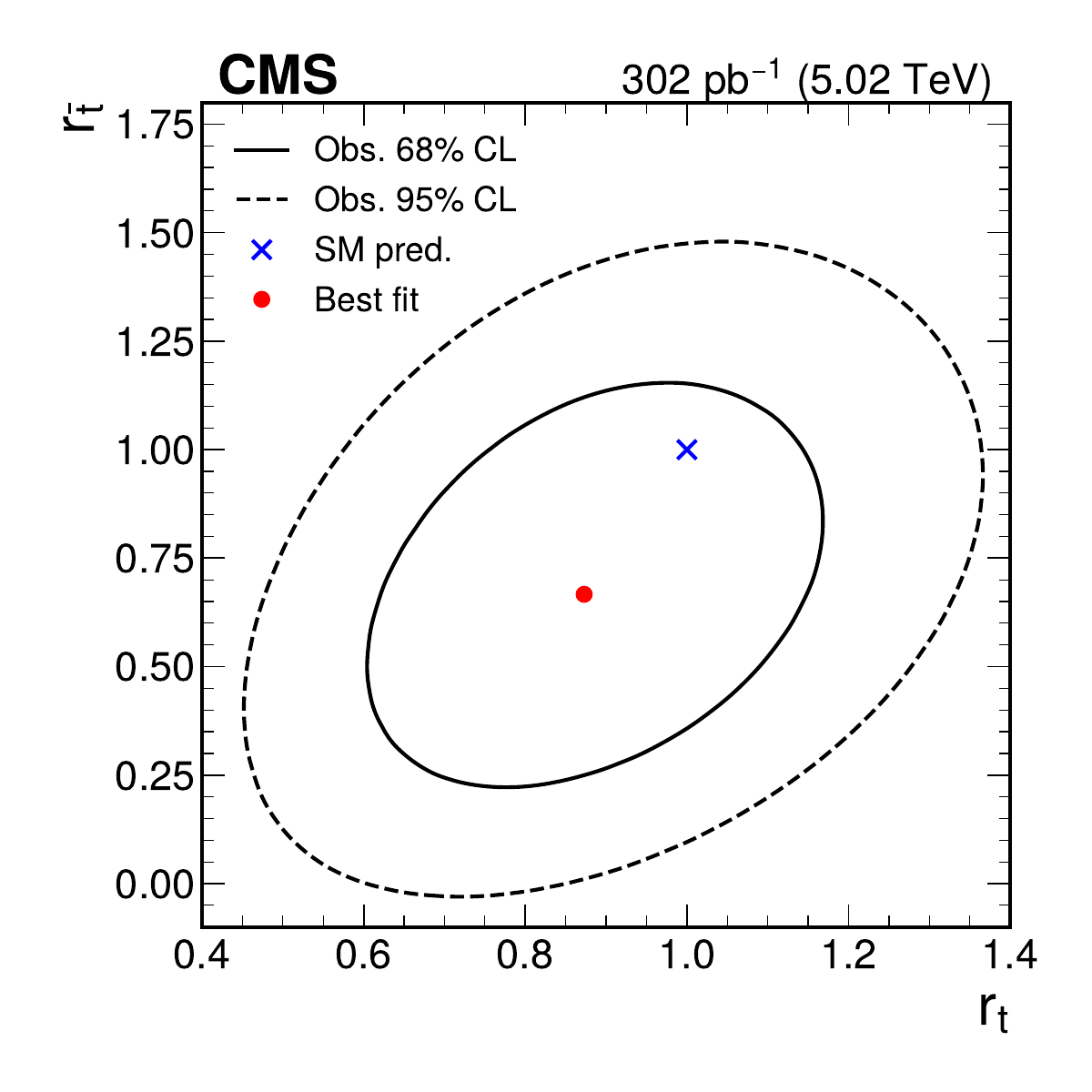}
\caption{Two-dimensional 68 and 95\% confidence level regions for the signal strengths modifying the top quark ($r_{\PQt}$) and antiquark ($r_{\PAQt}$) cross sections individually, represented by the solid and dashed lines respectively. The blue cross indicates the SM prediction and the red dot the best fit for both signal strengths.}
\label{fig:2dscan}
\end{figure}

Using a similar approach, likelihood scans are performed for the top quark and antiquark cross sections separately. In this case, the likelihood is minimized leaving both the signal strength for the top quark and for the top antiquark cross section unconstrained. The top quark cross section is measured to be $\sto=17.7\,^{+3.9}_{-3.7}\pb$, corresponding to a relative uncertainty of around $\pm$21\%, with a systematic uncertainty of $\pm$14\%, a luminosity uncertainty of $\pm$1.9\%, and a statistical uncertainty of $\pm$16\%.
The top antiquark cross section is measured to be $\santitop=6.7\,^{+3.2}_{-3.0}\pb$, corresponding to a relative uncertainty of around $\pm$46\%, with a systematic uncertainty of around $\pm$34\%, a luminosity uncertainty of $\pm$1.9\%, and a statistical uncertainty of around $\pm$30\%. The correlation coefficient between the two signal strengths is 0.36 and the two-dimensional confidence region for both parameters is shown in Fig.~\ref{fig:2dscan}, in which a good agreement with the SM predictions is seen.

Finally, the ratio of the top quark and antiquark cross sections is also extracted by performing a third ML fit to the same distributions, where the cross section ratio and the inclusive cross section are treated as unconstrained parameters.
This procedure permits to handle correlated uncertainties correctly, which is not achieved by taking the arithmetic ratio of the previously reported individual cross sections.
With this approach, the ratio is found to be $\ratio=2.7\,^{+1.5}_{-0.8}\stat\,^{+1.2}_{-0.3}\syst=2.7\,^{+1.9}_{-0.8}$.

In addition to these measurements, the $t$-channel total cross section can be used to calculate the absolute value of the Cabibbo--Kobayashi--Maskawa ME, $\abs{\Vtb}$, since it has a dependence on $\fLV^2\abs{\Vtb}^2$. In this context, \fLV is an anomalous form factor which is unity under the SM, or $\fLV\neq1$ for physics beyond the SM.
Under the assumption that the other MEs, $\abs{V_{\PQt\PQd}}$ and $\abs{V_{\PQt\PQs}}$, are significantly smaller than $\abs{\Vtb}$, and that the top quark decays only to a \PQb quark and a \PW boson, $\absfLVVtb=0.92\pm0.09\experr\pm0.01\thy$ is measured. The first uncertainty accounts for all the uncertainties in the measurement whereas the second one considers the uncertainties of the theoretical prediction.

\begin{figure}[!htp]
\centering
\includegraphics[width=0.6\textwidth]{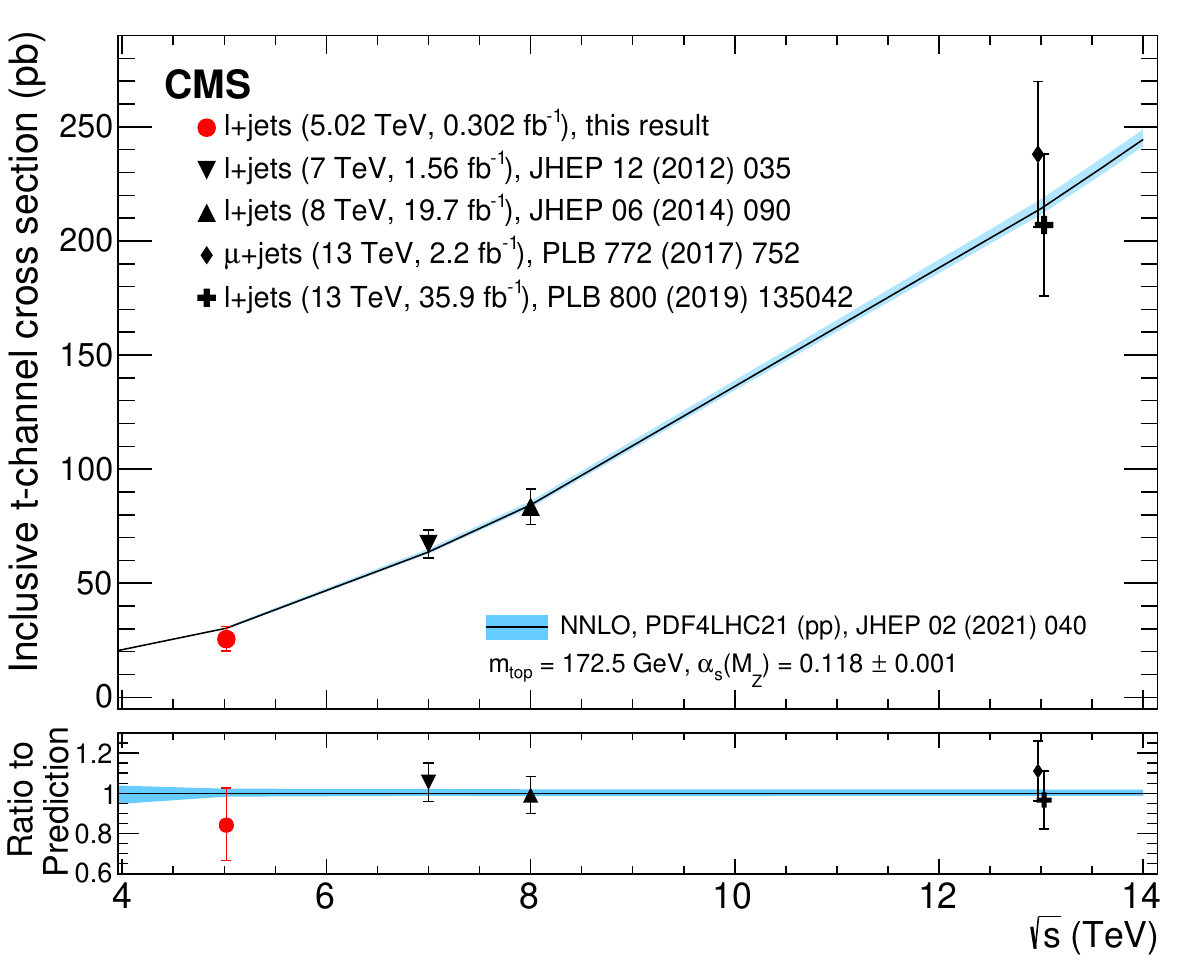}
\caption{Summary of the CMS measurements of the $t$-channel single top quark production cross section as a function of $\sqrt{s}$.}
\label{fig:summaryplots}
\end{figure}

\begin{figure}[!htp]
\centering
\includegraphics[width=0.48\textwidth]{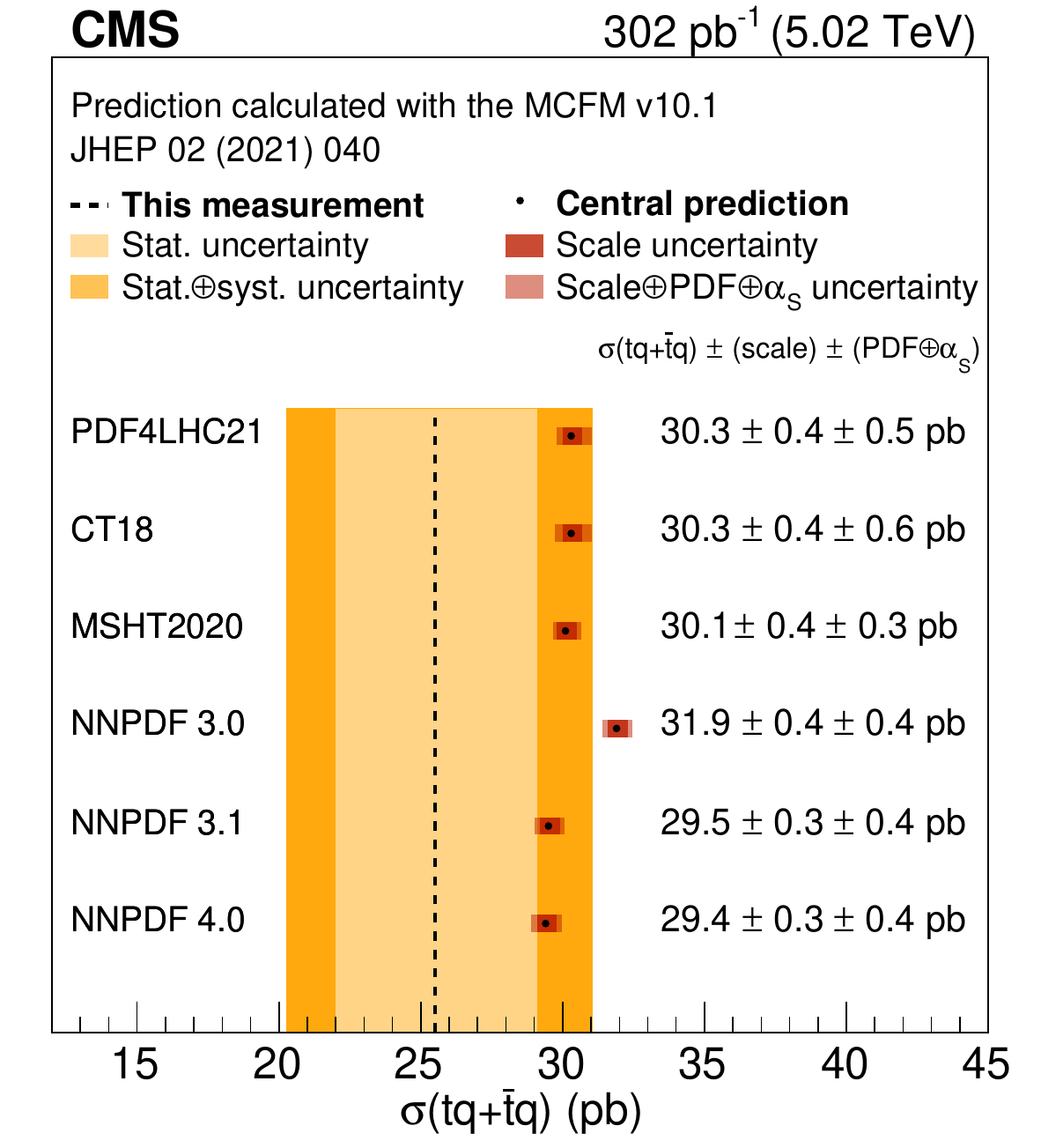}
\hfill
\includegraphics[width=0.48\textwidth]{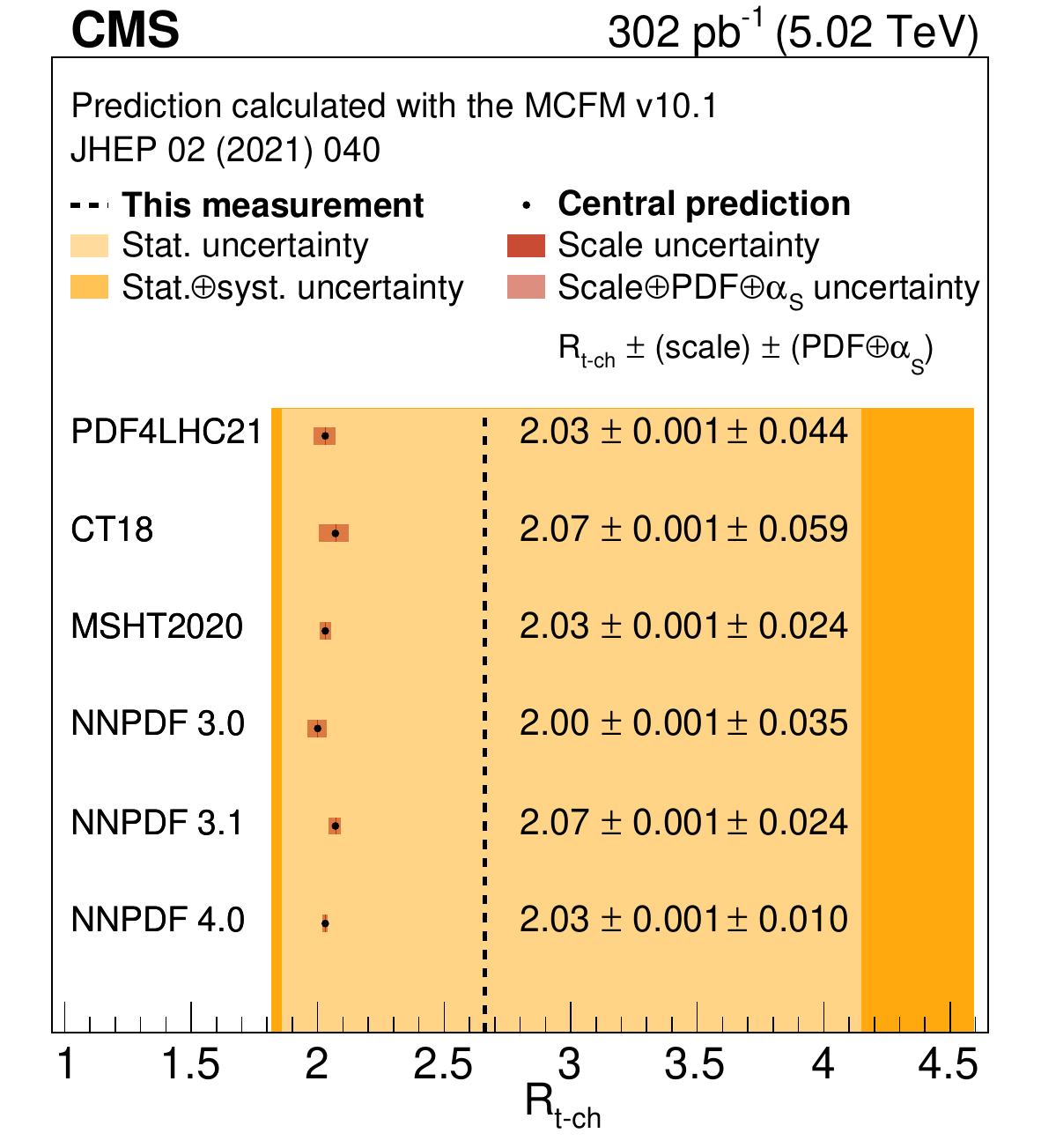}
\caption{Comparison of the measured values at $\sqrt{s}=5.02\TeV$ with predictions using different PDF sets, calculated with the \MCFM program~\cite{updatedxsecs} for \sboth (left) and \ratio (right).}
\label{fig:pdfcomparison}
\end{figure}

Figure~\ref{fig:summaryplots} presents a summary of $t$-channel single top quark cross section measurements by the CMS Collaboration at different $\sqrt{s}$, including the measurement described in this paper.
All of them show a good agreement with SM predictions. Also, results presented here are consistent with those obtained by the ATLAS experiment at the same energy~\cite{atlas5tev}.
In addition, a comparison between the NNLO prediction from the \MCFM MC generator with different PDF sets~\cite{updatedxsecs} for both the measured cross section and the ratio of invididual cross sections is shown in Fig.~\ref{fig:pdfcomparison}, where the expected results using different PDF sets are all seen to agree well with the measured values.

\section{Summary}
\label{sec:Summary}

A measurement of the single top quark $t$-channel production cross section in proton-proton collisions at a centre-of-mass energy of 5.02\TeV is performed
for events with one electron or one muon and multiple jets using data collected by the CMS experiment in 2017 in conditions of low number of additional interactions per bunch crossing, corresponding to an integrated luminosity of 302\pbinv. This represents the first CMS measurement of the process at that energy.
The main background in the analysis originates from \wjets and top quark pair production, both estimated using simulation. Smaller contributions from quantum chromodynamics multijet events are derived from data. The cross section is measured through a maximum likelihood fit to the observed event yields in twelve categories defined in terms of the number of jets and \PQb-tagged jets, the lepton flavour and electric charge.
The cross sections are found to be $\sboth=25.5\,^{+3.6}_{-3.5}\stat\,^{+4.2}_{-3.9}\syst\pm0.5\lum\pb$, $\sto=17.7\,^{+2.9}_{-2.7}\stat\,^{+2.6}_{-2.4}\syst\pm0.3\lum\pb$, and $\santitop=6.7\,^{+2.4}_{-1.6}\stat\,^{+2.1}_{-2.5}\syst\pm0.1\lum\pb$. The ratio between the individual cross sections is found to be $\ratio=2.7\,^{+1.5}_{-0.8}\stat\,^{+1.2}_{-0.3}\syst$.
These results are used to calculate the absolute value of the Cabibbo--Kobayashi--Maskawa matrix element, which is found to be $\absfLVVtb=0.92\pm0.09\experr\pm0.01\thy$.
All the results are in good agreement with the standard model predictions.

\begin{acknowledgments}
We congratulate our colleagues in the CERN accelerator departments for the excellent performance of the LHC and thank the technical and administrative staffs at CERN and at other CMS institutes for their contributions to the success of the CMS effort. In addition, we gratefully acknowledge the computing centres and personnel of the Worldwide LHC Computing Grid and other centres for delivering so effectively the computing infrastructure essential to our analyses. Finally, we acknowledge the enduring support for the construction and operation of the LHC, the CMS detector, and the supporting computing infrastructure provided by the following funding agencies: SC (Armenia), BMBWF and FWF (Austria); FNRS and FWO (Belgium); CNPq, CAPES, FAPERJ, FAPERGS, and FAPESP (Brazil); MES and BNSF (Bulgaria); CERN; CAS, MoST, and NSFC (China); MINCIENCIAS (Colombia); MSES and CSF (Croatia); RIF (Cyprus); SENESCYT (Ecuador); ERC PRG and PSG, TARISTU24-TK10 and MoER TK202 (Estonia); Academy of Finland, MEC, and HIP (Finland); CEA and CNRS/IN2P3 (France); SRNSF (Georgia); BMFTR, DFG, and HGF (Germany); GSRI (Greece); NKFIH (Hungary); DAE and DST (India); IPM (Iran); SFI (Ireland); INFN (Italy); MSIT and NRF (Republic of Korea); MES (Latvia); LMTLT (Lithuania); MOE and UM (Malaysia); BUAP, CINVESTAV, CONACYT, LNS, SEP, and UASLP-FAI (Mexico); MOS (Montenegro); MBIE (New Zealand); PAEC (Pakistan); MES, NSC, and NAWA (Poland); FCT (Portugal); MESTD (Serbia); MICIU/AEI and PCTI (Spain); MOSTR (Sri Lanka); Swiss Funding Agencies (Switzerland); MST (Taipei); MHESI (Thailand); TUBITAK and TENMAK (T\"{u}rkiye); NASU (Ukraine); STFC (United Kingdom); DOE and NSF (USA).

\begin{sloppypar}
\setlength\emergencystretch{\hsize}
\hyphenation{Rachada-pisek} Individuals have received support from the Marie-Curie programme and the European Research Council and Horizon 2020 Grant, contract Nos.\ 675440, 724704, 752730, 758316, 765710, 824093, 101115353, 101002207, 101001205, and COST Action CA16108 (European Union); the Leventis Foundation; the Alfred P.\ Sloan Foundation; the Alexander von Humboldt Foundation; the Science Committee, project no. 22rl-037 (Armenia); the Fonds pour la Formation \`a la Recherche dans l'Industrie et dans l'Agriculture (FRIA) and Fonds voor Wetenschappelijk Onderzoek contract No. 1228724N (Belgium); the Beijing Municipal Science \& Technology Commission, No. Z191100007219010, the Fundamental Research Funds for the Central Universities, the Ministry of Science and Technology of China under Grant No. 2023YFA1605804, the Natural Science Foundation of China under Grant No. 12535004, and USTC Research Funds of the Double First-Class Initiative No.\ YD2030002017 (China); the Ministry of Education, Youth and Sports (MEYS) of the Czech Republic; the Shota Rustaveli National Science Foundation, grant FR-22-985 (Georgia); the Deutsche Forschungsgemeinschaft (DFG), among others, under Germany's Excellence Strategy -- EXC 2121 ``Quantum Universe" -- 390833306, and under project number 400140256 - GRK2497; the Hellenic Foundation for Research and Innovation (HFRI), Project Number 2288 (Greece); the Hungarian Academy of Sciences, the New National Excellence Program - \'UNKP, the NKFIH research grants K 131991, K 133046, K 138136, K 143460, K 143477, K 146913, K 146914, K 147048, 2020-2.2.1-ED-2021-00181, TKP2021-NKTA-64, and 2025-1.1.5-NEMZ\_KI-2025-00004 (Hungary); the Council of Science and Industrial Research, India; ICSC -- National Research Centre for High Performance Computing, Big Data and Quantum Computing, FAIR -- Future Artificial Intelligence Research, and CUP I53D23001070006 (Mission 4 Component 1), funded by the NextGenerationEU program, the Italian Ministry of University and Research (MUR) under Bando PRIN 2022 -- CUP I53C24002390006, PRIN PRIMULA 2022RBYK7T (Italy); the Latvian Council of Science; the Ministry of Education and Science, project no. 2022/WK/14, and the National Science Center, contracts Opus 2021/41/B/ST2/01369, 2021/43/B/ST2/01552, 2023/49/B/ST2/03273, and the NAWA contract BPN/PPO/2021/1/00011 (Poland); the Funda\c{c}\~ao para a Ci\^encia e a Tecnologia (Portugal); the National Priorities Research Program by Qatar National Research Fund; MICIU/AEI/10.13039/501100011033, ERDF/EU, ``European Union NextGenerationEU/PRTR", projects PID2022-142604OB-C21, PID2022-139519OB-C21, PID2023-147706NB-I00, PID2023-148896NB-I00, PID2023-146983NB-I00, PID2023-147115NB-I00, PID2023-148418NB-C41, PID2023-148418NB-C42, PID2023-148418NB-C43, PID2023-148418NB-C44, PID2024-158190NB-C22, RYC2021-033305-I, RYC2024-048719-I, CNS2023-144781, CNS2024-154769 and Plan de Ciencia, Tecnolog{\'i}a e Innovaci{\'o}n de Asturias, Spain; the Chulalongkorn Academic into Its 2nd Century Project Advancement Project, the National Science, Research and Innovation Fund program IND\_FF\_68\_369\_2300\_097, and the Program Management Unit for Human Resources \& Institutional Development, Research and Innovation, grant B39G680009 (Thailand); the Eric \& Wendy Schmidt Fund for Strategic Innovation through the CERN Next Generation Triggers project under grant agreement number SIF-2023-004; the Kavli Foundation; the Nvidia Corporation; the SuperMicro Corporation; the Welch Foundation, contract C-1845; and the Weston Havens Foundation (USA).
\end{sloppypar}
\end{acknowledgments}

\bibliography{auto_generated}
\cleardoublepage \appendix\section{The CMS Collaboration \label{app:collab}}\begin{sloppypar}\hyphenpenalty=5000\widowpenalty=500\clubpenalty=5000\input{TOP-24-011-public-authorlist.tex}\end{sloppypar}
\end{document}

%% file: TOP-24-011-public-authorlist.tex
\cmsinstitute{Yerevan Physics Institute, Yerevan, Armenia}
{\tolerance=6000
A.~Hayrapetyan, V.~Makarenko\cmsorcid{0000-0002-8406-8605}, A.~Tumasyan\cmsAuthorMark{1}\cmsorcid{0009-0000-0684-6742}
\par}
\cmsinstitute{Institut f\"{u}r Hochenergiephysik, Vienna, Austria}
{\tolerance=6000
W.~Adam\cmsorcid{0000-0001-9099-4341}, L.~Benato\cmsorcid{0000-0001-5135-7489}, T.~Bergauer\cmsorcid{0000-0002-5786-0293}, M.~Dragicevic\cmsorcid{0000-0003-1967-6783}, C.~Giordano\cmsorcid{0000-0001-6317-2481}, P.S.~Hussain\cmsorcid{0000-0002-4825-5278}, M.~Jeitler\cmsAuthorMark{2}\cmsorcid{0000-0002-5141-9560}, N.~Krammer\cmsorcid{0000-0002-0548-0985}, A.~Li\cmsorcid{0000-0002-4547-116X}, D.~Liko\cmsorcid{0000-0002-3380-473X}, M.~Matthewman, J.~Schieck\cmsAuthorMark{2}\cmsorcid{0000-0002-1058-8093}, R.~Sch\"{o}fbeck\cmsAuthorMark{2}\cmsorcid{0000-0002-2332-8784}, M.~Shooshtari\cmsorcid{0009-0004-8882-4887}, M.~Sonawane\cmsorcid{0000-0003-0510-7010}, W.~Waltenberger\cmsorcid{0000-0002-6215-7228}, C.-E.~Wulz\cmsAuthorMark{2}\cmsorcid{0000-0001-9226-5812}
% ...
\par}
\vskip\cmsinstskip
\dag:~Deceased\\
$^{1}$Also at Yerevan State University, Yerevan, Armenia\\
$^{2}$Also at TU Wien, Vienna, Austria\\
% ...


%% file: TOP-24-011_temp.bbl
\providecommand{\href}[2]{#2}\begingroup\raggedright\begin{thebibliography}{10}%
\makeatletter
\providecommand{\hrefCMSnoop }[0]{\@secondoftwo}%
\makeatother
\providecommand{\doi}{\texttt{doi:}\begingroup \urlstyle{tt}\Url}

\bibitem{Bigi:1986jk}
I.~I.~Y. Bigi\hrefCMSnoop {}{ { et~al.}, ``{Production and Decay Properties of
  Ultraheavy Quarks}'',} \textit{ Phys. Lett. B} \textbf{ 181} (1986) 157,
  \href{http://dx.doi.org/10.1016/0370-2693(86)91275-X}{\doi{10.1016/0370-2693(86)91275-X}}.

\bibitem{theo:Alwall_2006}
J.~Alwall\hrefCMSnoop {}{ { et~al.}, ``Is ${V}_{\PQt\PQb}\simeq1$?'',} \textit{
  Eur. Phys. J. C} \textbf{ 49} (2007) 791,
  \href{http://dx.doi.org/10.1140/epjc/s10052-006-0137-y}{\doi{10.1140/epjc/s10052-006-0137-y}},
  \href{http://www.arXiv.org/abs/hep-ph/0607115}{\texttt{arXiv:hep-ph/0607115}}.

\bibitem{theo:Holdom_2009}
B.~Holdom\hrefCMSnoop {}{ { et~al.}, ``Four statements about the fourth
  generation'',} \textit{ PMC Phys. A} \textbf{ 3} (2009) 4,
  \href{http://dx.doi.org/10.1186/1754-0410-3-4}{\doi{10.1186/1754-0410-3-4}},
  \href{http://www.arXiv.org/abs/0904.4698}{\texttt{arXiv:0904.4698}}.

\bibitem{Tevatron}
\hrefCMSnoop {}{{\DZERO} Collaboration, ``Model-independent measurement of
  $t$-channel single top quark production in ${\Pp\PAp}$ collisions at
  $\sqrt{s}={1.96\TeV}$'',} \textit{ Phys. Lett. B} \textbf{ 705} (2011) 313,
  \href{http://dx.doi.org/10.1016/j.physletb.2011.10.035}{\doi{10.1016/j.physletb.2011.10.035}},
  \href{http://www.arXiv.org/abs/1105.2788}{\texttt{arXiv:1105.2788}}.

\bibitem{Chatrchyan:2011vp}
\hrefCMSnoop {}{{CMS Collaboration}, ``Measurement of the $t$-channel single
  top quark production cross section in ${\Pp\Pp}$ collisions at
  $\sqrt{s}={7\TeV}$'',} \textit{ Phys. Rev. Lett.} \textbf{ 107} (2011)
  091802,
  \href{http://dx.doi.org/10.1103/PhysRevLett.107.091802}{\doi{10.1103/PhysRevLett.107.091802}},
  \href{http://www.arXiv.org/abs/1106.3052}{\texttt{arXiv:1106.3052}}.

\bibitem{Chatrchyan:2012ep}
\hrefCMSnoop {}{{CMS Collaboration}, ``Measurement of the single-top-quark
  $t$-channel cross section in ${\Pp\Pp}$ collisions at $\sqrt{s}={7\TeV}$'',}
  \textit{ JHEP} \textbf{ 12} (2012) 035,
  \href{http://dx.doi.org/10.1007/JHEP12(2012)035}{\doi{10.1007/JHEP12(2012)035}},
  \href{http://www.arXiv.org/abs/1209.4533}{\texttt{arXiv:1209.4533}}.

\bibitem{atlas7}
\hrefCMSnoop {}{{ATLAS Collaboration}, ``Comprehensive measurements of
  $t$-channel single top-quark production cross sections at $\sqrt{s}={7\TeV}$
  with the {ATLAS} detector'',} \textit{ Phys. Rev. D} \textbf{ 90} (2014)
  112006,
  \href{http://dx.doi.org/10.1103/PhysRevD.90.112006}{\doi{10.1103/PhysRevD.90.112006}},
  \href{http://www.arXiv.org/abs/1406.7844}{\texttt{arXiv:1406.7844}}.

\bibitem{Khachatryan:2014iya}
\hrefCMSnoop {}{{CMS Collaboration}, ``Measurement of the $t$-channel
  single-top-quark production cross section and of the $\abs{V_{\PQt\PQb}}$
  {CKM} matrix element in ${\Pp\Pp}$ collisions at $\sqrt{s}={8\TeV}$'',}
  \textit{ JHEP} \textbf{ 06} (2014) 090,
  \href{http://dx.doi.org/10.1007/JHEP06(2014)090}{\doi{10.1007/JHEP06(2014)090}},
  \href{http://www.arXiv.org/abs/1403.7366}{\texttt{arXiv:1403.7366}}.

\bibitem{atlas8}
\hrefCMSnoop {}{{ATLAS Collaboration}, ``Fiducial, total and differential
  cross-section measurements of $t$-channel single top-quark production in
  ${\Pp\Pp}$ collisions at {8\TeV} using data collected by the {ATLAS}
  detector'',} \textit{ Eur. Phys. J. C} \textbf{ 77} (2017) 531,
  \href{http://dx.doi.org/10.1140/epjc/s10052-017-5061-9}{\doi{10.1140/epjc/s10052-017-5061-9}},
  \href{http://www.arXiv.org/abs/1702.02859}{\texttt{arXiv:1702.02859}}.

\bibitem{Sirunyan:2016cdg}
\hrefCMSnoop {}{{CMS Collaboration}, ``Cross section measurement of $t$-channel
  single top quark production in ${\Pp\Pp}$ collisions at
  $\sqrt{s}={13\TeV}$'',} \textit{ Phys. Lett. B} \textbf{ 772} (2017) 752,
  \href{http://dx.doi.org/10.1016/j.physletb.2017.07.047}{\doi{10.1016/j.physletb.2017.07.047}},
  \href{http://www.arXiv.org/abs/1610.00678}{\texttt{arXiv:1610.00678}}.

\bibitem{top201711}
\hrefCMSnoop {}{{CMS Collaboration}, ``Measurement of the single top quark and
  antiquark production cross sections in the $t$ channel and their ratio in
  proton-proton collisions at $\sqrt{s}={13\TeV}$'',} \textit{ Phys. Lett. B}
  \textbf{ 800} (2020) 135042,
  \href{http://dx.doi.org/10.1016/j.physletb.2019.135042}{\doi{10.1016/j.physletb.2019.135042}},
  \href{http://www.arXiv.org/abs/1812.10514}{\texttt{arXiv:1812.10514}}.

\bibitem{atlas13}
\hrefCMSnoop {}{{ATLAS Collaboration}, ``Measurement of $t$-channel production
  of single top quarks and antiquarks in ${\Pp\Pp}$ collisions at {13\TeV}
  using the full {ATLAS} \mbox{Run 2} data sample'',} \textit{ JHEP} \textbf{
  05} (2024) 305,
  \href{http://dx.doi.org/10.1007/JHEP05(2024)305}{\doi{10.1007/JHEP05(2024)305}},
  \href{http://www.arXiv.org/abs/2403.02126}{\texttt{arXiv:2403.02126}}.

\bibitem{ATLAS:2026vxj}
\hrefCMSnoop {}{{ATLAS Collaboration}, ``{Measurement of differential
  $t$-channel single top (anti)quark production cross-sections at 13 TeV with
  the ATLAS detector}'',} 2026.
  \href{http://www.arXiv.org/abs/2601.04938}{\texttt{arXiv:2601.04938}}.
  Submitted to \textit{JHEP.}

\bibitem{atlas5tev}
\hrefCMSnoop {}{{ATLAS Collaboration}, ``Measurement of $t$-channel
  single-top-quark production in ${\Pp\Pp}$ collisions at $\sqrt{s}={5.02\TeV}$
  with the {ATLAS} detector'',} \textit{ Phys. Lett. B} \textbf{ 854} (2024)
  138726,
  \href{http://dx.doi.org/10.1016/j.physletb.2024.138726}{\doi{10.1016/j.physletb.2024.138726}},
  \href{http://www.arXiv.org/abs/2310.01518}{\texttt{arXiv:2310.01518}}.

\bibitem{wz5tev}
\hrefCMSnoop {}{{CMS Collaboration}, ``Measurements of the electroweak diboson
  production cross sections in proton-proton collisions at
  $\sqrt{s}={5.02\TeV}$ using leptonic decays'',} \textit{ Phys. Rev. Lett.}
  \textbf{ 127} (2021) 191801,
  \href{http://dx.doi.org/10.1103/PhysRevLett.127.191801}{\doi{10.1103/PhysRevLett.127.191801}},
  \href{http://www.arXiv.org/abs/2107.01137}{\texttt{arXiv:2107.01137}}.

\bibitem{CMS:2021gwv}
\hrefCMSnoop {}{{CMS Collaboration}, ``Measurement of the inclusive \ttbar
  production cross section in proton-proton collisions at
  $\sqrt{s}={5.02\TeV}$'',} \textit{ JHEP} \textbf{ 04} (2022) 144,
  \href{http://dx.doi.org/10.1007/JHEP04(2022)144}{\doi{10.1007/JHEP04(2022)144}},
  \href{http://www.arXiv.org/abs/2112.09114}{\texttt{arXiv:2112.09114}}.

\bibitem{CMS:2024ghc}
\hrefCMSnoop {}{{CMS Collaboration}, ``Measurement of the inclusive \ttbar
  cross section in final states with at least one lepton and additional jets
  with 302\pbinv of ${\Pp\Pp}$ collisions at $\sqrt{s}={5.02\TeV}$'',} \textit{
  JHEP} \textbf{ 04} (2025) 099,
  \href{http://dx.doi.org/10.1007/JHEP04(2025)099}{\doi{10.1007/JHEP04(2025)099}},
  \href{http://www.arXiv.org/abs/2410.21631}{\texttt{arXiv:2410.21631}}.

\bibitem{hepdata}
\hrefCMSnoop {}{}{HEPData} record for this analysis, 2026.
\newblock
  \href{http://dx.doi.org/10.17182/hepdata.161623}{\doi{10.17182/hepdata.161623}}.

\bibitem{CMS:2008zzk}
\hrefCMSnoop {}{{CMS Collaboration}, ``The {CMS} experiment at the {CERN}
  {LHC}'',} \textit{ JINST} \textbf{ 3} (2008) S08004,
  \href{http://dx.doi.org/10.1088/1748-0221/3/08/S08004}{\doi{10.1088/1748-0221/3/08/S08004}}.

\bibitem{CMS:2023gfb}
\hrefCMSnoop {}{{CMS Collaboration}, ``Development of the {CMS} detector for
  the {CERN} {LHC} \mbox{Run 3}'',} \textit{ JINST} \textbf{ 19} (2024) P05064,
  \href{http://dx.doi.org/10.1088/1748-0221/19/05/P05064}{\doi{10.1088/1748-0221/19/05/P05064}},
  \href{http://www.arXiv.org/abs/2309.05466}{\texttt{arXiv:2309.05466}}.

\bibitem{Sirunyan:2020zal}
\hrefCMSnoop {}{{CMS Collaboration}, ``Performance of the {CMS} {\Lone} trigger
  in proton-proton collisions at $\sqrt{s}={13\TeV}$'',} \textit{ JINST}
  \textbf{ 15} (2020) P10017,
  \href{http://dx.doi.org/10.1088/1748-0221/15/10/P10017}{\doi{10.1088/1748-0221/15/10/P10017}},
  \href{http://www.arXiv.org/abs/2006.10165}{\texttt{arXiv:2006.10165}}.

\bibitem{Khachatryan:2016bia}
\hrefCMSnoop {}{{CMS Collaboration}, ``The {CMS} trigger system'',} \textit{
  JINST} \textbf{ 12} (2017) P01020,
  \href{http://dx.doi.org/10.1088/1748-0221/12/01/P01020}{\doi{10.1088/1748-0221/12/01/P01020}},
  \href{http://www.arXiv.org/abs/1609.02366}{\texttt{arXiv:1609.02366}}.

\bibitem{CMS:2024aqx}
\hrefCMSnoop {}{{CMS Collaboration}, ``Performance of the {CMS} high-level
  trigger during {LHC} \mbox{Run 2}'',} \textit{ JINST} \textbf{ 19} (2024)
  P11021,
  \href{http://dx.doi.org/10.1088/1748-0221/19/11/P11021}{\doi{10.1088/1748-0221/19/11/P11021}},
  \href{http://www.arXiv.org/abs/2410.17038}{\texttt{arXiv:2410.17038}}.

\bibitem{CMS:2020uim}
\hrefCMSnoop {}{{CMS Collaboration}, ``Electron and photon reconstruction and
  identification with the {CMS} experiment at the {CERN} {LHC}'',} \textit{
  JINST} \textbf{ 16} (2021) P05014,
  \href{http://dx.doi.org/10.1088/1748-0221/16/05/P05014}{\doi{10.1088/1748-0221/16/05/P05014}},
  \href{http://www.arXiv.org/abs/2012.06888}{\texttt{arXiv:2012.06888}}.

\bibitem{Sirunyan:2018fpa}
\hrefCMSnoop {}{{CMS Collaboration}, ``Performance of the {CMS} muon detector
  and muon reconstruction with proton-proton collisions at
  $\sqrt{s}={13\TeV}$'',} \textit{ JINST} \textbf{ 13} (2018) P06015,
  \href{http://dx.doi.org/10.1088/1748-0221/13/06/P06015}{\doi{10.1088/1748-0221/13/06/P06015}},
  \href{http://www.arXiv.org/abs/1804.04528}{\texttt{arXiv:1804.04528}}.

\bibitem{CMS:2014pgm}
\hrefCMSnoop {}{{CMS Collaboration}, ``Description and performance of track and
  primary-vertex reconstruction with the {CMS} tracker'',} \textit{ JINST}
  \textbf{ 9} (2014) P10009,
  \href{http://dx.doi.org/10.1088/1748-0221/9/10/P10009}{\doi{10.1088/1748-0221/9/10/P10009}},
  \href{http://www.arXiv.org/abs/1405.6569}{\texttt{arXiv:1405.6569}}.

\bibitem{CMS-PRF-14-001}
\hrefCMSnoop {}{{CMS Collaboration}, ``Particle-flow reconstruction and global
  event description with the {CMS} detector'',} \textit{ JINST} \textbf{ 12}
  (2017) P10003,
  \href{http://dx.doi.org/10.1088/1748-0221/12/10/P10003}{\doi{10.1088/1748-0221/12/10/P10003}},
  \href{http://www.arXiv.org/abs/1706.04965}{\texttt{arXiv:1706.04965}}.

\bibitem{ptmissperformance}
\hrefCMSnoop {}{{CMS Collaboration}, ``Performance of missing transverse
  momentum reconstruction in proton-proton collisions at $\sqrt{s}={13\TeV}$
  using the {CMS} detector'',} \textit{ JINST} \textbf{ 14} (2019) P07004,
  \href{http://dx.doi.org/10.1088/1748-0221/14/07/P07004}{\doi{10.1088/1748-0221/14/07/P07004}},
  \href{http://www.arXiv.org/abs/1903.06078}{\texttt{arXiv:1903.06078}}.

\bibitem{Cacciari:2008gp}
\hrefCMSnoop {}{M.~Cacciari, G.~P. Salam, and G.~Soyez, ``The anti-\kt jet
  clustering algorithm'',} \textit{ JHEP} \textbf{ 04} (2008) 063,
  \href{http://dx.doi.org/10.1088/1126-6708/2008/04/063}{\doi{10.1088/1126-6708/2008/04/063}},
  \href{http://www.arXiv.org/abs/0802.1189}{\texttt{arXiv:0802.1189}}.

\bibitem{Cacciari:2011ma}
\hrefCMSnoop {}{M.~Cacciari, G.~P. Salam, and G.~Soyez, ``{\FASTJET} user
  manual'',} \textit{ Eur. Phys. J. C} \textbf{ 72} (2012) 1896,
  \href{http://dx.doi.org/10.1140/epjc/s10052-012-1896-2}{\doi{10.1140/epjc/s10052-012-1896-2}},
  \href{http://www.arXiv.org/abs/1111.6097}{\texttt{arXiv:1111.6097}}.

\bibitem{Khachatryan:2016kdb}
\hrefCMSnoop {}{{CMS Collaboration}, ``Jet energy scale and resolution in the
  {CMS} experiment in ${\Pp\Pp}$ collisions at {8\TeV}'',} \textit{ JINST}
  \textbf{ 12} (2017) P02014,
  \href{http://dx.doi.org/10.1088/1748-0221/12/02/P02014}{\doi{10.1088/1748-0221/12/02/P02014}},
  \href{http://www.arXiv.org/abs/1607.03663}{\texttt{arXiv:1607.03663}}.

\bibitem{CMS:2017wyc}
\href {https://cds.cern.ch/record/2256875}{{CMS Collaboration}, ``Jet
  algorithms performance in {13\TeV} data'',} CMS Physics Analysis Summary
  CMS-PAS-JME-16-003, 2017.

\bibitem{Sirunyan:2017ezt}
\hrefCMSnoop {}{{CMS Collaboration}, ``Identification of heavy-flavour jets
  with the {CMS} detector in ${\Pp\Pp}$ collisions at {13\TeV}'',} \textit{
  JINST} \textbf{ 13} (2018) P05011,
  \href{http://dx.doi.org/10.1088/1748-0221/13/05/P05011}{\doi{10.1088/1748-0221/13/05/P05011}},
  \href{http://www.arXiv.org/abs/1712.07158}{\texttt{arXiv:1712.07158}}.

\bibitem{CMS-TDR-15-02}
\hrefCMSnoop {}{{CMS Collaboration}, ``Technical proposal for the {Phase-II}
  upgrade of the {Compact Muon Solenoid}'',} CMS Technical Proposal
  CERN-LHCC-2015-010, CMS-TDR-15-02, 2015.
\newblock
  \href{http://dx.doi.org/10.17181/CERN.VU8I.D59J}{\doi{10.17181/CERN.VU8I.D59J}}.

\bibitem{muonMVA}
\hrefCMSnoop {}{{CMS Collaboration}, ``Muon identification using multivariate
  techniques in the {CMS} experiment in proton-proton collisions at
  $\sqrt{s}={13\TeV}$'',} \textit{ JINST} \textbf{ 19} (2024) P02031,
  \href{http://dx.doi.org/10.1088/1748-0221/19/02/P02031}{\doi{10.1088/1748-0221/19/02/P02031}},
  \href{http://www.arXiv.org/abs/2310.03844}{\texttt{arXiv:2310.03844}}.

\bibitem{CMS:2020mpn}
\hrefCMSnoop {}{{CMS Collaboration}, ``Measurement of the {Higgs} boson
  production rate in association with top quarks in final states with
  electrons, muons, and hadronically decaying tau leptons at
  $\sqrt{s}={13\TeV}$'',} \textit{ Eur. Phys. J. C} \textbf{ 81} (2021) 378,
  \href{http://dx.doi.org/10.1140/epjc/s10052-021-09014-x}{\doi{10.1140/epjc/s10052-021-09014-x}},
  \href{http://www.arXiv.org/abs/2011.03652}{\texttt{arXiv:2011.03652}}.

\bibitem{Agostinelli:2002hh}
\hrefCMSnoop {}{{GEANT4} Collaboration, ``{\GEANTfour}---a simulation
  toolkit'',} \textit{ Nucl. Instrum. Meth. A} \textbf{ 506} (2003) 250,
  \href{http://dx.doi.org/10.1016/S0168-9002(03)01368-8}{\doi{10.1016/S0168-9002(03)01368-8}}.

\bibitem{Frixione:2007vw}
\hrefCMSnoop {}{S.~Frixione, P.~Nason, and C.~Oleari, ``Matching {NLO} {QCD}
  computations with parton shower simulations: the {\POWHEG} method'',}
  \textit{ JHEP} \textbf{ 11} (2007) 070,
  \href{http://dx.doi.org/10.1088/1126-6708/2007/11/070}{\doi{10.1088/1126-6708/2007/11/070}},
  \href{http://www.arXiv.org/abs/0709.2092}{\texttt{arXiv:0709.2092}}.

\bibitem{Alioli:2010xd}
\hrefCMSnoop {}{S.~Alioli, P.~Nason, C.~Oleari, and E.~Re, ``A general
  framework for implementing {NLO} calculations in shower {Monte Carlo}
  programs: the {\POWHEG} \textsc{box}'',} \textit{ JHEP} \textbf{ 06} (2010)
  043,
  \href{http://dx.doi.org/10.1007/JHEP06(2010)043}{\doi{10.1007/JHEP06(2010)043}},
  \href{http://www.arXiv.org/abs/1002.2581}{\texttt{arXiv:1002.2581}}.

\bibitem{Frixione:2007nw}
\hrefCMSnoop {}{S.~Frixione, G.~Ridolfi, and P.~Nason, ``A positive-weight
  next-to-leading-order {Monte Carlo} for heavy flavour hadroproduction'',}
  \textit{ JHEP} \textbf{ 09} (2007) 126,
  \href{http://dx.doi.org/10.1088/1126-6708/2007/09/126}{\doi{10.1088/1126-6708/2007/09/126}},
  \href{http://www.arXiv.org/abs/0707.3088}{\texttt{arXiv:0707.3088}}.

\bibitem{Nason:2004rx}
\hrefCMSnoop {}{P.~Nason, ``A new method for combining {NLO} {QCD} with shower
  {Monte Carlo} algorithms'',} \textit{ JHEP} \textbf{ 11} (2004) 040,
  \href{http://dx.doi.org/10.1088/1126-6708/2004/11/040}{\doi{10.1088/1126-6708/2004/11/040}},
  \href{http://www.arXiv.org/abs/hep-ph/0409146}{\texttt{arXiv:hep-ph/0409146}}.

\bibitem{Ball:2017nwa}
\hrefCMSnoop {}{{NNPDF} Collaboration, ``Parton distributions from
  high-precision collider data'',} \textit{ Eur. Phys. J. C} \textbf{ 77}
  (2017) 663,
  \href{http://dx.doi.org/10.1140/epjc/s10052-017-5199-5}{\doi{10.1140/epjc/s10052-017-5199-5}},
  \href{http://www.arXiv.org/abs/1706.00428}{\texttt{arXiv:1706.00428}}.

\bibitem{tWDR}
S.~Frixione\hrefCMSnoop {}{ { et~al.}, ``{Single-top hadroproduction in
  association with a W boson}'',} \textit{ JHEP} \textbf{ 07} (2008) 029,
  \href{http://dx.doi.org/10.1088/1126-6708/2008/07/029}{\doi{10.1088/1126-6708/2008/07/029}},
  \href{http://www.arXiv.org/abs/0805.3067}{\texttt{arXiv:0805.3067}}.

\bibitem{Alwall:2014hca}
J.~Alwall\hrefCMSnoop {}{ { et~al.}, ``The automated computation of tree-level
  and next-to-leading order differential cross sections, and their matching to
  parton shower simulations'',} \textit{ JHEP} \textbf{ 07} (2014) 079,
  \href{http://dx.doi.org/10.1007/JHEP07(2014)079}{\doi{10.1007/JHEP07(2014)079}},
  \href{http://www.arXiv.org/abs/1405.0301}{\texttt{arXiv:1405.0301}}.

\bibitem{Frederix:2012ps}
\hrefCMSnoop {}{R.~Frederix and S.~Frixione, ``Merging meets matching in
  {\MCATNLO}'',} \textit{ JHEP} \textbf{ 12} (2012) 061,
  \href{http://dx.doi.org/10.1007/JHEP12(2012)061}{\doi{10.1007/JHEP12(2012)061}},
  \href{http://www.arXiv.org/abs/1209.6215}{\texttt{arXiv:1209.6215}}.

\bibitem{Sjostrand:2014zea}
T.~Sj{\"o}strand\hrefCMSnoop {}{ { et~al.}, ``An introduction to
  {\PYTHIA8.2}'',} \textit{ Comput. Phys. Commun.} \textbf{ 191} (2015) 159,
  \href{http://dx.doi.org/10.1016/j.cpc.2015.01.024}{\doi{10.1016/j.cpc.2015.01.024}},
  \href{http://www.arXiv.org/abs/1410.3012}{\texttt{arXiv:1410.3012}}.

\bibitem{Sirunyan:2019dfx}
\hrefCMSnoop {}{{CMS Collaboration}, ``Extraction and validation of a new set
  of {CMS} {\PYTHIA8} tunes from underlying-event measurements'',} \textit{
  Eur. Phys. J. C} \textbf{ 80} (2020) 4,
  \href{http://dx.doi.org/10.1140/epjc/s10052-019-7499-4}{\doi{10.1140/epjc/s10052-019-7499-4}},
  \href{http://www.arXiv.org/abs/1903.12179}{\texttt{arXiv:1903.12179}}.

\bibitem{Campbell:2020fhf}
\hrefCMSnoop {}{J.~Campbell, T.~Neumann, and Z.~Sullivan, ``Single-top-quark
  production in the $t$-channel at {NNLO}'',} \textit{ JHEP} \textbf{ 02}
  (2021) 040,
  \href{http://dx.doi.org/10.1007/JHEP02(2021)040}{\doi{10.1007/JHEP02(2021)040}},
  \href{http://www.arXiv.org/abs/2012.01574}{\texttt{arXiv:2012.01574}}.

\bibitem{PDF4LHCWorkingGroup:2022cjn}
\hrefCMSnoop {}{{PDF4LHC Working Group}, R.~D. Ball { et~al.}, ``The
  {PDF4LHC21} combination of global {PDF} fits for the {LHC} \mbox{Run 3}'',}
  \textit{ J. Phys. G} \textbf{ 49} (2022) 080501,
  \href{http://dx.doi.org/10.1088/1361-6471/ac7216}{\doi{10.1088/1361-6471/ac7216}},
  \href{http://www.arXiv.org/abs/2203.05506}{\texttt{arXiv:2203.05506}}.

\bibitem{Czakon:2011xx}
\hrefCMSnoop {}{M.~Czakon and A.~Mitov, ``\textsc{top++}: a program for the
  calculation of the top-pair cross-section at hadron colliders'',} \textit{
  Comput. Phys. Commun.} \textbf{ 185} (2014) 2930,
  \href{http://dx.doi.org/10.1016/j.cpc.2014.06.021}{\doi{10.1016/j.cpc.2014.06.021}},
  \href{http://www.arXiv.org/abs/1112.5675}{\texttt{arXiv:1112.5675}}.

\bibitem{Czakon:2013goa}
\hrefCMSnoop {}{M.~Czakon, P.~Fiedler, and A.~Mitov, ``Total top-quark
  pair-production cross section at hadron colliders through
  $\mathcal{O}({\alpS}^4)$'',} \textit{ Phys. Rev. Lett.} \textbf{ 110} (2013)
  252004,
  \href{http://dx.doi.org/10.1103/PhysRevLett.110.252004}{\doi{10.1103/PhysRevLett.110.252004}},
  \href{http://www.arXiv.org/abs/1303.6254}{\texttt{arXiv:1303.6254}}.

\bibitem{Kidonakis:2021vob}
\hrefCMSnoop {}{N.~Kidonakis and N.~Yamanaka, ``Higher-order corrections for
  ${\PQt\PW}$ production at high-energy hadron colliders'',} \textit{ JHEP}
  \textbf{ 05} (2021) 278,
  \href{http://dx.doi.org/10.1007/JHEP05(2021)278}{\doi{10.1007/JHEP05(2021)278}},
  \href{http://www.arXiv.org/abs/2102.11300}{\texttt{arXiv:2102.11300}}.

\bibitem{Melnikov:2006kv}
\hrefCMSnoop {}{K.~Melnikov and F.~Petriello, ``Electroweak gauge boson
  production at hadron colliders through $\mathcal{O}({\alpS^2})$'',} \textit{
  Phys. Rev. D} \textbf{ 74} (2006) 114017,
  \href{http://dx.doi.org/10.1103/PhysRevD.74.114017}{\doi{10.1103/PhysRevD.74.114017}},
  \href{http://www.arXiv.org/abs/hep-ph/0609070}{\texttt{arXiv:hep-ph/0609070}}.

\bibitem{updatedxsecs}
\href {https://cds.cern.ch/record/2942746}{{ATLAS Collaboration}, ``Reference
  single top-quark cross-sections for {ATLAS} and {CMS} analyses'',} ATLAS PUB
  Note ATL-PHYS-PUB-2025-035, 2025.

\bibitem{cutler2012random}
\hrefCMSnoop {}{L.~Breiman, ``Random forests'',} \textit{ Mach. Learn.}
  \textbf{ 45} (2001) 5,
  \href{http://dx.doi.org/10.1023/A:1010933404324}{\doi{10.1023/A:1010933404324}}.

\bibitem{Conway:2011in}
\hrefCMSnoop {}{J.~S. Conway, ``Incorporating nuisance parameters in
  likelihoods for multisource spectra'',} in \textit{ {Proc. 2011 Workshop on
  Statistical Issues Related to Discovery Claims in Search Experiments and
  Unfolding (PHYSTAT 2011): Geneva, Switzerland, January 17--20, 2011}}.
\newblock 2011.
\newblock \href{http://www.arXiv.org/abs/1103.0354}{\texttt{arXiv:1103.0354}}.
\newblock
  \href{http://dx.doi.org/10.5170/CERN-2011-006.115}{\doi{10.5170/CERN-2011-006.115}}.

\bibitem{CMS:2010svw}
\hrefCMSnoop {}{{CMS Collaboration}, ``Measurements of inclusive {\PW} and
  {\PZ} cross sections in ${\Pp\Pp}$ collisions at $\sqrt{s}={7\TeV}$'',}
  \textit{ JHEP} \textbf{ 01} (2011) 080,
  \href{http://dx.doi.org/10.1007/JHEP01(2011)080}{\doi{10.1007/JHEP01(2011)080}},
  \href{http://www.arXiv.org/abs/1012.2466}{\texttt{arXiv:1012.2466}}.

\bibitem{clopper}
\hrefCMSnoop {}{C.~J. Clopper and E.~S. Pearson, ``The use of confidence or
  fiducial limits illustrated in the case of the binomial'',} \textit{
  Biometrika} \textbf{ 26} (1934) 404,
  \href{http://dx.doi.org/10.1093/biomet/26.4.404}{\doi{10.1093/biomet/26.4.404}}.

\bibitem{lowess}
\hrefCMSnoop {}{W.~S. Cleveland, ``Robust locally weighted regression and
  smoothing scatterplots'',} \textit{ J. Am. Stat. Assoc.} \textbf{ 74} (1979)
  829, \href{http://dx.doi.org/10.2307/2286407}{\doi{10.2307/2286407}}.

\bibitem{Butterworth:2015oua}
\hrefCMSnoop {}{J.~Butterworth { et~al.}, ``{PDF4LHC} recommendations for {LHC}
  \mbox{Run 2}'',} \textit{ J. Phys. G} \textbf{ 43} (2016) 023001,
  \href{http://dx.doi.org/10.1088/0954-3899/43/2/023001}{\doi{10.1088/0954-3899/43/2/023001}},
  \href{http://www.arXiv.org/abs/1510.03865}{\texttt{arXiv:1510.03865}}.

\bibitem{Sirunyan:2017ule}
\hrefCMSnoop {}{{CMS Collaboration}, ``Measurement of the inclusive \ttbar
  cross section in ${\Pp\Pp}$ collisions at $\sqrt{s}={5.02\TeV}$ using final
  states with at least one charged lepton'',} \textit{ JHEP} \textbf{ 03}
  (2018) 115,
  \href{http://dx.doi.org/10.1007/JHEP03(2018)115}{\doi{10.1007/JHEP03(2018)115}},
  \href{http://www.arXiv.org/abs/1711.03143}{\texttt{arXiv:1711.03143}}.

\bibitem{wjetsref}
\hrefCMSnoop {}{{CMS Collaboration}, ``Measurement of the production cross
  section of a {\PW} boson in association with two {\PQb} jets in ${\Pp\Pp}$
  collisions at $\sqrt{s}={8\TeV}$'',} \textit{ Eur. Phys. J. C} \textbf{ 77}
  (2017) 92,
  \href{http://dx.doi.org/10.1140/epjc/s10052-016-4573-z}{\doi{10.1140/epjc/s10052-016-4573-z}},
  \href{http://www.arXiv.org/abs/1608.07561}{\texttt{arXiv:1608.07561}}.

\bibitem{bib:LUM-19-001}
\href {https://cds.cern.ch/record/2765655}{{CMS Collaboration}, ``Luminosity
  measurement in proton-proton collisions at {5.02\TeV} in 2017 at {CMS}'',}
  CMS Physics Analysis Summary CMS-PAS-LUM-19-001, 2021.

\bibitem{CMS:LUM-17-003}
\hrefCMSnoop {}{{CMS Collaboration}, ``Precision luminosity measurement in
  proton-proton collisions at $\sqrt{s}={13\TeV}$ in 2015 and 2016 at {CMS}'',}
  \textit{ Eur. Phys. J. C} \textbf{ 81} (2021) 800,
  \href{http://dx.doi.org/10.1140/epjc/s10052-021-09538-2}{\doi{10.1140/epjc/s10052-021-09538-2}},
  \href{http://www.arXiv.org/abs/2104.01927}{\texttt{arXiv:2104.01927}}.

\bibitem{cmscollaboration2024cms}
\hrefCMSnoop {}{{CMS Collaboration}, ``The {CMS} statistical analysis and
  combination tool: \textsc{combine}'',} \textit{ Comput. Softw. Big Sci.}
  \textbf{ 8} (2024) 19,
  \href{http://dx.doi.org/10.1007/s41781-024-00121-4}{\doi{10.1007/s41781-024-00121-4}},
  \href{http://www.arXiv.org/abs/2404.06614}{\texttt{arXiv:2404.06614}}.

\end{thebibliography}\endgroup
